\documentclass[12pt]{emulateapj}
\slugcomment{Accepted, to appear in ApJ, v710, 2010 February}
\pdfoutput=1
\usepackage{wrapfig}
\usepackage{graphicx}
\usepackage{amssymb}
\usepackage{amsmath}
\usepackage{ifpdf}
\usepackage{longtable}
\usepackage{rotating}

\begin{document}

\newcommand\etal{{\it et al.}}
\newcommand\cf{{\it cf.~}}
\newcommand\eg{{\it e.g.,~}}
\newcommand\ie{{\it i.e.,~}}

\title{Three-dimensional Simulations of Bi-directed Magnetohydrodynamic Jets Interacting with Cluster Environments}

\author{S. M. O'Neill\altaffilmark{1}, T. W. Jones\altaffilmark{2}}
\altaffiltext{1}{University of Maryland, Department of Astronomy and Maryland Astronomy Center for Theory and Computation, College Park, MD, 20742; soneill@astro.umd.edu}
\altaffiltext{2}{School of Physics and Astronomy, University of Minnesota, Minneapolis, MN 55455; twj@msi.umn.edu}

\begin{abstract}
We report on a series of three-dimensional magnetohydrodynamic simulations of active galactic nucleus (AGN) jet propagation in realistic models of magnetized galaxy clusters.
We are primarily interested in the details of energy transfer between jets and the intracluster medium (ICM) to help clarify what role such flows could have in the reheating of cluster cores.
Our simulated jets feature a range of intermittency behaviors, including intermittent jets that periodically switch on and off and one model jet that shuts down completely, naturally creating a relic plume.
The ICM into which these jets propagate incorporates tangled magnetic field geometries and density substructure designed to mimic some likely features of real galaxy clusters.
We find that our jets are characteristically at least $60\%$ efficient at transferring thermal energy to the ICM.
Irreversible heat energy is not uniformly distributed, however, instead residing preferentially in regions very near the jet/cocoon boundaries.
While intermittency affects the details of how, when, and where this energy is deposited, all of our models generically fail to heat the cluster cores uniformly.
Both the detailed density structure and nominally weak magnetic fields in the ICM play interesting roles in perturbing the flows, particularly when the jets are non-steady.
Still, this perturbation is never sufficient to isotropize the jet energy deposition, suggesting that some other ingredient is required for AGN jets to successfully reheat cluster cores. 
\end{abstract}


\section{Introduction}\label{sec:intro}
Over the past decade, X-ray observations of galaxy clusters have clearly illustrated the immense energies associated with interactions between active galactic nuclei (AGNs) and their environments.
For example, discoveries of X-ray cavities associated with AGN radio-emitting `relic bubbles' ({\it e.g.} \citealt{mcnamaraetal00, fabianetal00,fujitaetal02,nulsenetal02,wiseetal07,sandersetal09}) demonstrate that radio lobes associated with current or previous AGN activity often completely displace the surrounding intracluster medium (ICM).
Likewise, X-ray observations of `ghost cavities' devoid of radio emission ({\it e.g.} \citealt{mcnamaraetal01, mazzottaetal02}) have shown that these cavities maintain some structural integrity even after the radio population has aged beyond the limits of detectability.
The energies associated with these plasma bubbles are found to be upward of $10^{59}-10^{60}~$erg \citep{birzanetal04,dunnetal05,wiseetal07}, suggesting that they could play a significant role in stifling cooling flows and maintaining the $\sim 2~$keV temperature floor observed in clusters ({\it e.g.} \citealt{petersonetal01,fabianetal01,kaastraetal01,tamuraetal01}).
This idea is further supported by simulations of cluster formation and evolution with AGN feedback, such as those conducted by \citet{bruggenetal05} and \citet{sijackispringel06}. 

Observations of detached relic bubbles and evolved ghost cavities also suggest that a complete picture of AGN interactions with the ICM should incorporate a history of realistic jet lifetimes and duty cycles. 
Some attempts to estimate AGN intermittency are based largely on energetics.
\citet{sokeretal01}, for example, estimate jet lifetimes of $10^7-10^8~$years with duty cycles of $5 \%$ or less to stifle cooling while matching optical and radio observations of cooling cores.
Observations of bubbles in the absence of AGN \citep{mazzottaetal02} have also been used to estimate lifetimes of $10^7-10^8~$years and duty cycles of at most $10 \%$.
Other approaches focus more on timescales associated with observed features in individual sources, such as those conducted by \citet{fabianetal03}, \citet{formanetal05}, \citet{jamrozyetal07}, and \citet{blantonetal09}, all of whom find a rate of one event per $10^7-10^8~$years.
Observations of the so-called double-double radio sources \citep{schoenmakersetal00}, in which two distinct episodes of jet activity are clearly visible, have implied rather short AGN lifetimes of $\sim 10^6~$years at most.
\citet{croometal04} take a completely different approach, using QSO clustering to estimate AGN intermittency, and their result suggests AGN lifetimes of $10^6-10^7~$years.
Most recently, \citet{birzanetal09} use a complete sample of clusters to estimate duty cycles of at least $60\%$ for AGN outbursts in cooling flow clusters.
\citet{diehletal08} also examine a large sample of clusters and infer that the observed size distribution of X-ray cavities is most consistent with either continuous hydrodynamic inflation or discrete production of bubbles by magnetically dominated jets.

Although there is much observational evidence to support models of intermittent jets that feed energetic plasma bubbles to the ICM, much still needs to be understood about the precise nature of energy flow in these systems.
Simple analytic models, such as those developed by \citet{cioffiblondin92}, \citet{kaiseralexander97}, and \citet{komissarovfalle98}, are powerful tools for analyzing the basic dynamics and energetics in these systems, but are not designed to address the details of energy transfer to the ICM.
Furthermore, these models tend to feature arguments based upon self-similarity or constant jet luminosities, and so may not apply to realistic ICM atmospheres or intermittent jets.
Likewise, numerical studies that focused on jet dynamics in both the non-relativistic ({\it e.g.} \citealt{normanetal82, hardeeclarke92,norman96,carvalhoodea02a,carvalhoodea02b,krause03,krause05}) and relativistic regimes ({\it e.g.} \citealt{duncanhughes94, martietal97, aloyetal99, mizunoetal07, keppensetal08, mignoneetal09}) made great strides toward understanding how jets propagate, but were not necessarily interested in the details of energy transfer to the ICM.
\citet{rosenetal99} provide a useful analysis of the primary differences between relativistic and non-relativistic simulations, pointing out that jets of equal thrust have similar ages and efficiencies in both cases, although relativistic jets typically develop smaller cocoons than their non-relativistic counterparts.

Recent numerical studies of jet energetics, on the other hand, have contributed greatly to the understanding of energy flow from AGN to their environments.
\citet{zannietal05}, for example, conducted an ensemble of two-dimensional simulations of jets in detailed cluster environments, finding that a significant fraction of jet energy is deposited irreversibly into the ICM.
Likewise, our previous study of three-dimensional magnetohydrodynamic (MHD) jets \citep{oneilletal05} found that energy transfer to the ICM was efficient for both uniform and stratified environments.
Continuing work in three dimensions, \citet{vernaleoreynolds06} pointed out that simple hydrodynamic models of jet feedback characteristically develop a relatively lossless low-density channel and subsequently fail to dissipate and distribute energy widely enough to prevent catastrophic cooling.
\citet{heinzetal06}, however, attempted to address this issue by simulating wobbling hydrodynamic jets in an evolving cluster atmosphere that served to isotropize the deposited energy.
Whether the jet and jet-blown cocoon remain characteristically stable or break apart or mix with the ambient medium (as illustrated recently in \citealt{gaibleretal09}, for example) clearly influences the efficiency with which jets deliver energy to cluster cores.

Some recent numerical studies of relic bubble propagation also have had implications for energy transfer and transport in the ICM.
Simulations conducted by \citet{robinsonetal04} and \citet{jonesdeyoung05} in two-dimensions and \citet{ruszkowskietal07}, \citet{oneilletal09}, and \citet{dongstone09} in three dimensions, for example, illustrated that rising MHD bubbles lift a great deal of ICM material in their wake.
This material has the potential to mix and distribute energy within the ICM.
A recent paper by \citet{bruggenetal09} points out that this mixing at the bubble interface is further enhanced in the hydrodynamic case when turbulence is treated with a subgrid approach, which explicitly models the energetics and sizes of turbulent eddies that are not properly resolved by the numerical grid.
This is in contrast to the usual method of treating turbulence in which the computational grid resolution sets the effective scale of dissipation in the system.
It is currently unclear how this approach will affect the case of magnetized bubbles.
\citet{heinzchurazov05}, taking a different approach, showed that hydrodynamic bubbles could explicitly transform wave energy into heat that could be dissipated.

Several groups have also explored the effects of jet intermittency on flow dynamics and energetics.
The earliest example of a simulated restarting jet was conducted by \citet{clarkeburns91}, who found that subsequent jets moved quickly through previous jet material, forming a new bow shock in the process.
Using a two-dimensional hydrodynamics code, \citet{reynoldsetal02} found that plumes from dead radio galaxies lifted substantial amounts of ICM material as they rose and that a large fraction of the total jet energy went into heating the ICM.
This was followed by \citet{bassonalexander03}, who demonstrated in three dimensions that a buoyant jet-blown bubble continues to dredge up material from the cluster core for much longer than the jet lifetime. 
Also in three dimensions, \citet{ommaetal04} found that most of the jet energy in an outburst ends up too far away from the cluster core.
In a subsequent simulation that included cooling, \citet{ommabinney04} found that the amount of dense, cool gas immediately above the central AGN had a pronounced effect on where most of the jet energy was dissipated.
While the previously mentioned work by \citet{vernaleoreynolds06} showed that this effect was never sufficient to completely offset cooling in a simple ICM atmosphere, recent simulations by \citet{bruggen09} suggest that the modeling of subgrid turbulence may alleviate some of these problems.

Our study combines several of these different investigations, exploring the interactions of intermittent jets with the ICM.
We seek to determine how intermittency affects energy flow, asking in what form, where, and how rapidly energy is transferred from jet-produced structures to the ICM.
These simulations specifically examine whether a combination of jet intermittency, magnetic stresses, and ambient density fluctuations alone can lead to broader distribution and dissipation of jet energy, or whether something like the more chaotic jet model of \citet{heinzetal06}, or a very high Reynolds number turbulence such as that proposed by \citet{bruggen09}, is required.
Furthermore, we seek to understand the basic dynamics and morphology of intermittent flows.
We test how flow structures produced by intermittent and extinct jets evolve, how they transport and transfer energy, and how they interact with and potentially enhance ambient magnetic fields.

To address these questions, we employ a set of MHD simulations of jets in realistic cluster environments.
The jets are oppositely directed, launching into an ICM with full three-dimensional gravity, tangled magnetic fields, and density fluctuations designed to mimic cluster substructure.
Section \ref{sec:calc} describes the details of the numerical methods and the models employed.
Section \ref{sec:disc} contains a discussion of the results and astrophysical implications, while Section \ref{sec:conc} lists the main conclusions of this work.

\section{Calculation Details}\label{sec:calc}
Our simulations employ a second-order Eulerian total variation diminishing (TVD) non-relativistic ideal MHD code, described in \citet{ryuj95} and \citet{ryuetal98}.
The code explicitly enforces the divergence-free condition for magnetic fields through a constrained transport scheme detailed in \citet{ryuetal98}.
The numerical method conserves mass, momentum, and energy to machine accuracy.
Gravitational energy is handled in a fully conservative fashion by including this component in the total energy of the gas.
Additionally, energy fluxes across the boundaries and changes in energy on the grid are tracked and output for each energy type to facilitate analysis of energy flow (see Section \ref{sec:energy}).
Model parameters (discussed in Section \ref{sec:ambient}) are selected such that the cluster cooling times are longer than the simulation times ($t_{\rm cool} > 250$ Myr in the cluster cores), allowing us to neglect cooling in our simulations.

All simulations described here are performed on a three-dimensional Cartesian grid of full physical dimensions $x = 600 {\rm~kpc},~y = z = 480 {\rm~kpc}$.
Each zone represents one cubic kiloparsec of volume with $\Delta x = \Delta y = \Delta z = 1$ kpc.
Given that the total jet diameter is $14$ kpc (see Section 2.1), this is sufficient to resolve shocks and multiple scales of turbulence within the jet and cocoon.
With the exception of one model (AM, described in Section \ref{sec:jets}), these simulations feature sets of oppositely directed jets emanating from a cylindrical internal boundary located in the central region of the computational grid ($x = y = z = 0$).
The jet region is updated prior to each TVD step, and the settings in this region are used to control jet luminosity and intermittency.

\begin{deluxetable}{crrr}
\label{table}
\tabletypesize{\scriptsize}
\tablecaption{Summary of Simulations of Double Jets in Galaxy Clusters}
\tablewidth{0pt}
\tablehead{
\colhead{ID \tablenotemark{1}} &
\colhead{Jet Intermittency \tablenotemark{2}} &
\colhead{Final Age} \\}
\startdata
ST & Steady inflow for duration & $\approx 59 {\rm~Myr}$\\
I13 & Switch on/off every 13 Myr & $\approx 173 {\rm~Myr}$\\
I26 & Switch on/off every 26 Myr & $\approx 104 {\rm~Myr}$\\
RE & On for 26 Myr, then off & $\approx 144 {\rm~Myr}$\\
AM & No jets included & $\approx 183 {\rm~Myr}$\\
\enddata
\tablenotetext{1}{All models feature identical computational grids ($600 {\rm~kpc} \times 480 {\rm~kpc} \times 480 {\rm~kpc}$) and ICM structures.  ICM model features core density $\rho_0 = 8.33 \times 10^{-26}$ g cm$^{-3}$ and pressure $P_0 = 4.0 \times 10^{-10}~{\rm dyne~cm}^{-2}$, corresponding to a core sound speed $c_{0} = 895 {\rm~km~s^{-1}}$ and temperature $2.5 {\rm~keV}$.  Ambient magnetic fields are tangled with a characteristic core value $B_0 \sim 10~\mu {\rm G}$.}
\tablenotetext{2}{In fully active states, jets feature Mach number $M_j = 30$, corresponding to velocity $v_j = 0.0895c$.  Jet densities are calculated from fixed $\rho_{j} = \eta~\rho_0$, with $\eta = 0.01$.  Jet magnetic fields are completely toroidal with maximum strength $B_j = 10~\mu {\rm G}$.}
\end{deluxetable}

Since the jets are launched from the center of the computational grid, the outer grid boundary conditions have little influence on the evolution of jet/cocoon structures.
Still, the outer boundaries are crucial in maintaining the dynamical stability of the constantly evolving ambient medium.
In the outer grid boundaries, we use a set of modified continuous boundaries designed to maintain approximate hydrostatic equilibrium in the ambient medium while minimizing the impact of small-amplitude waves incident upon the boundaries.
Specifically, we apply the constraints that the sound speed be constant and that hydrostatic equilibrium apply to set the density and pressure conditions across a given boundary. 
We find that these boundaries maintain the total mass and energy of the grid to an accuracy greater than $99.5\%$ in the absence of a jet over three grid sound-crossing times, a time much longer than any of our jet simulations.

A passive color tracer $C_j$, representing the mass fraction of jet material, is introduced through jet flow to identify material that has passed through the jet orifice.
Additionally, we include a population of passive, nonthermal, relativistic cosmic ray particles in our models.
These cosmic rays are transported, injected, accelerated, and aged in a self-consistent fashion to enable the construction of realistic synthetic observations (see, {\it e.g.}, \citealt{jonesetal99}, \citealt{tregillisetal01a}, \citealt{tregillisetal04}, \citealt{mendygraletal10}).

Here, we describe five simulations in which the initial ambient conditions are precisely the same, but the time histories of jet activity vary.
In the following two subsections, we describe the jets and ambient media in detail.
The physical parameters of each simulation are described in Table 1.

\begin{figure*}[t]
\begin{center}
\includegraphics[type=pdf,ext=.pdf,read=.pdf,width=0.45\textwidth]{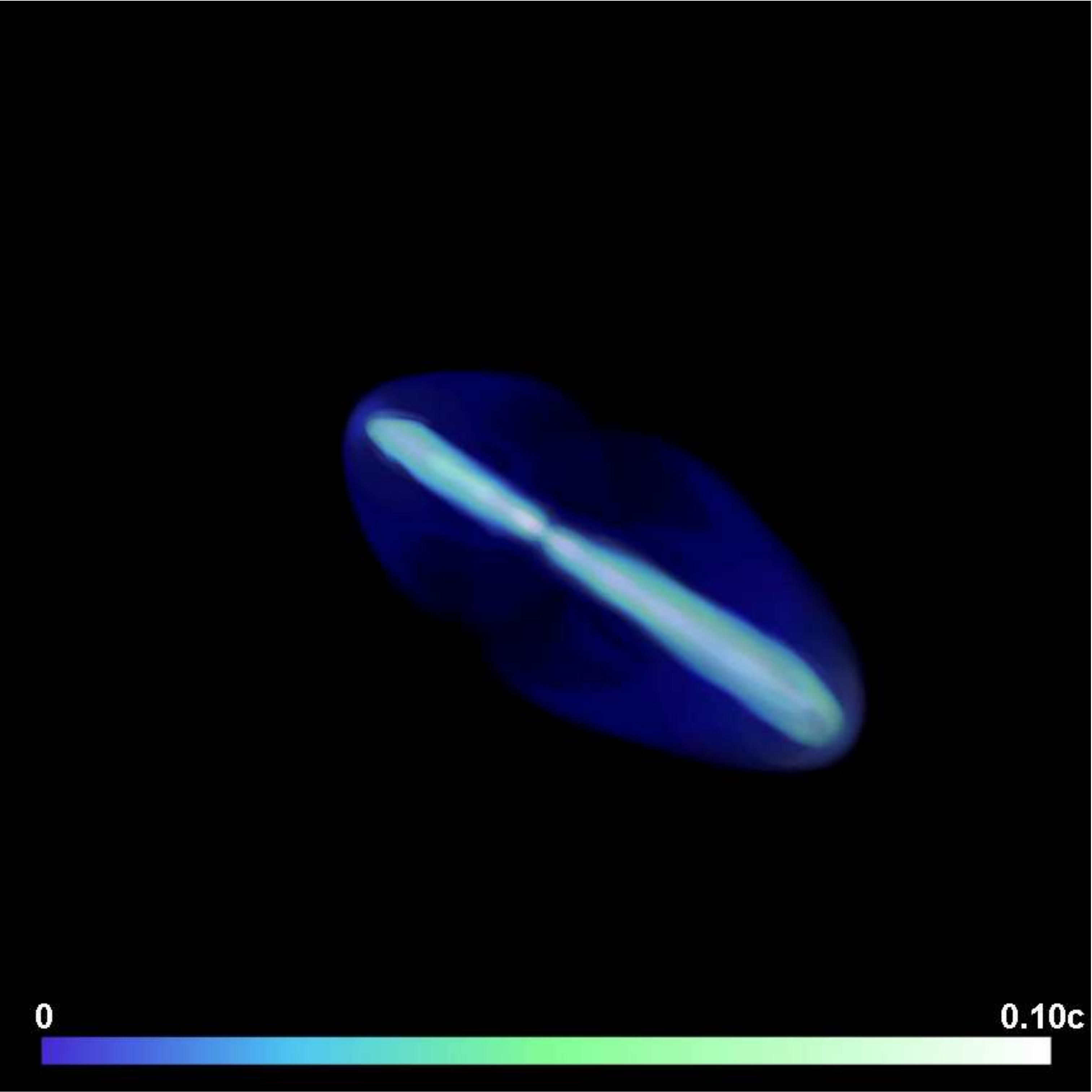}
\includegraphics[type=pdf,ext=.pdf,read=.pdf,width=0.45\textwidth]{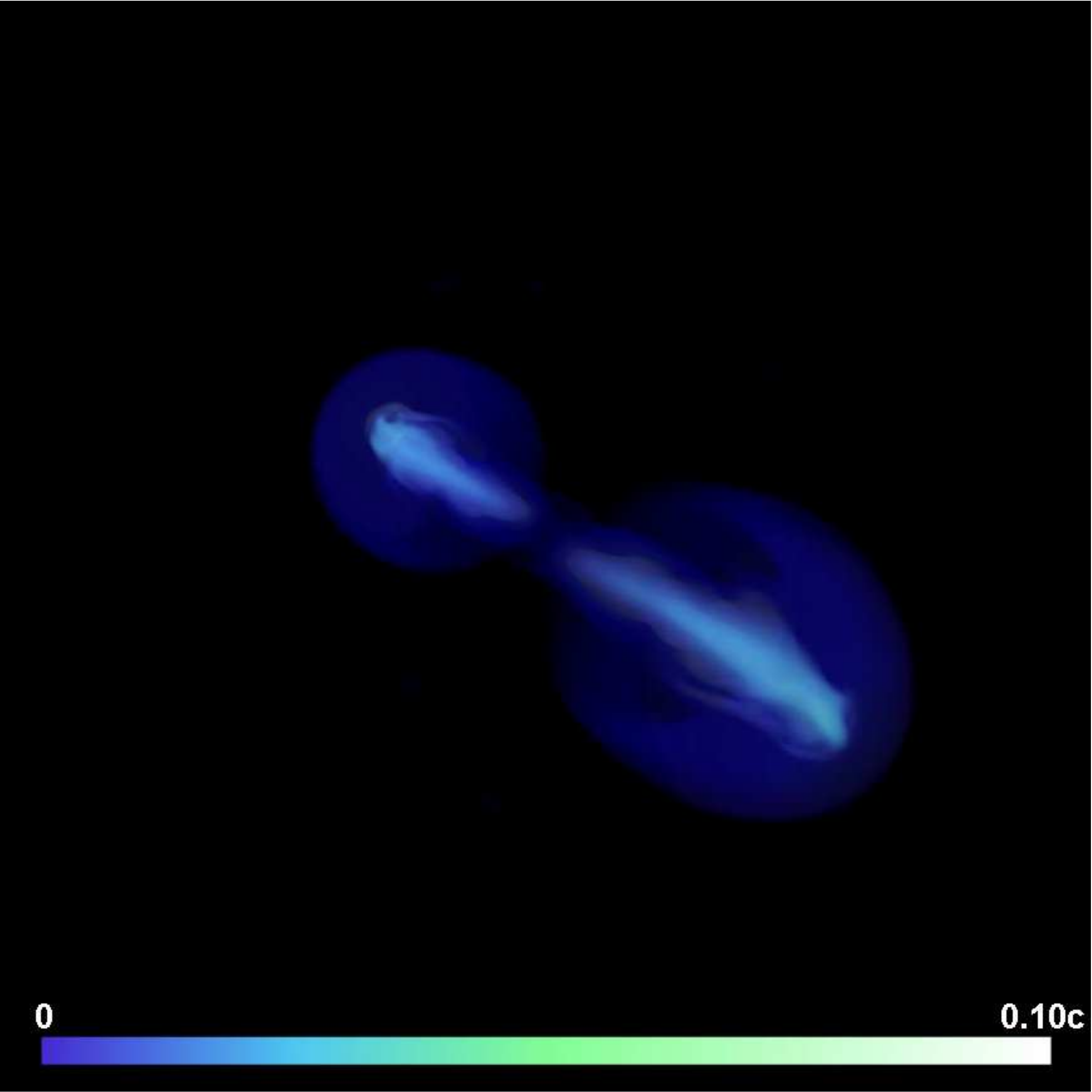}
\\[0.75em]
\includegraphics[type=pdf,ext=.pdf,read=.pdf,width=0.45\textwidth]{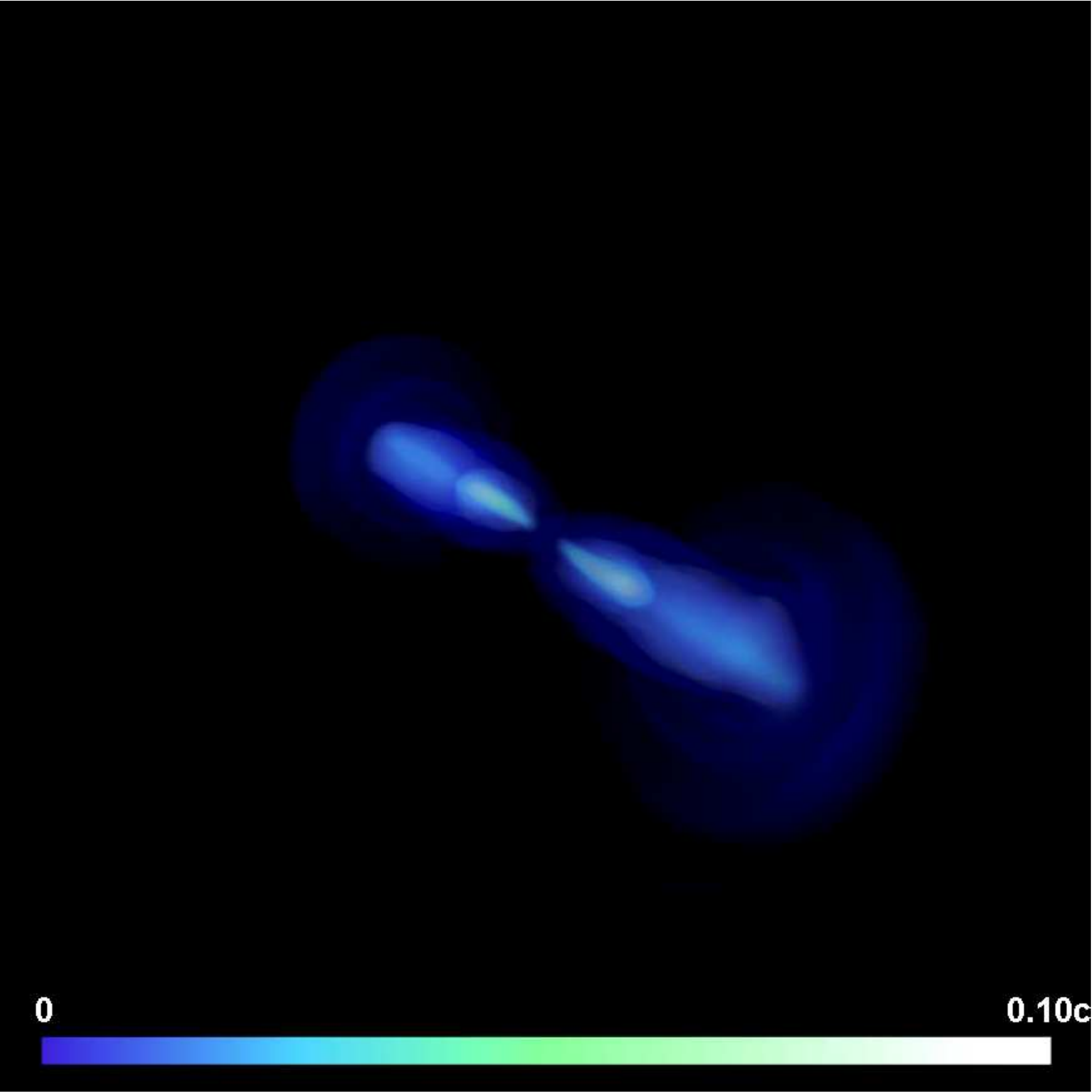}
\includegraphics[type=pdf,ext=.pdf,read=.pdf,width=0.45\textwidth]{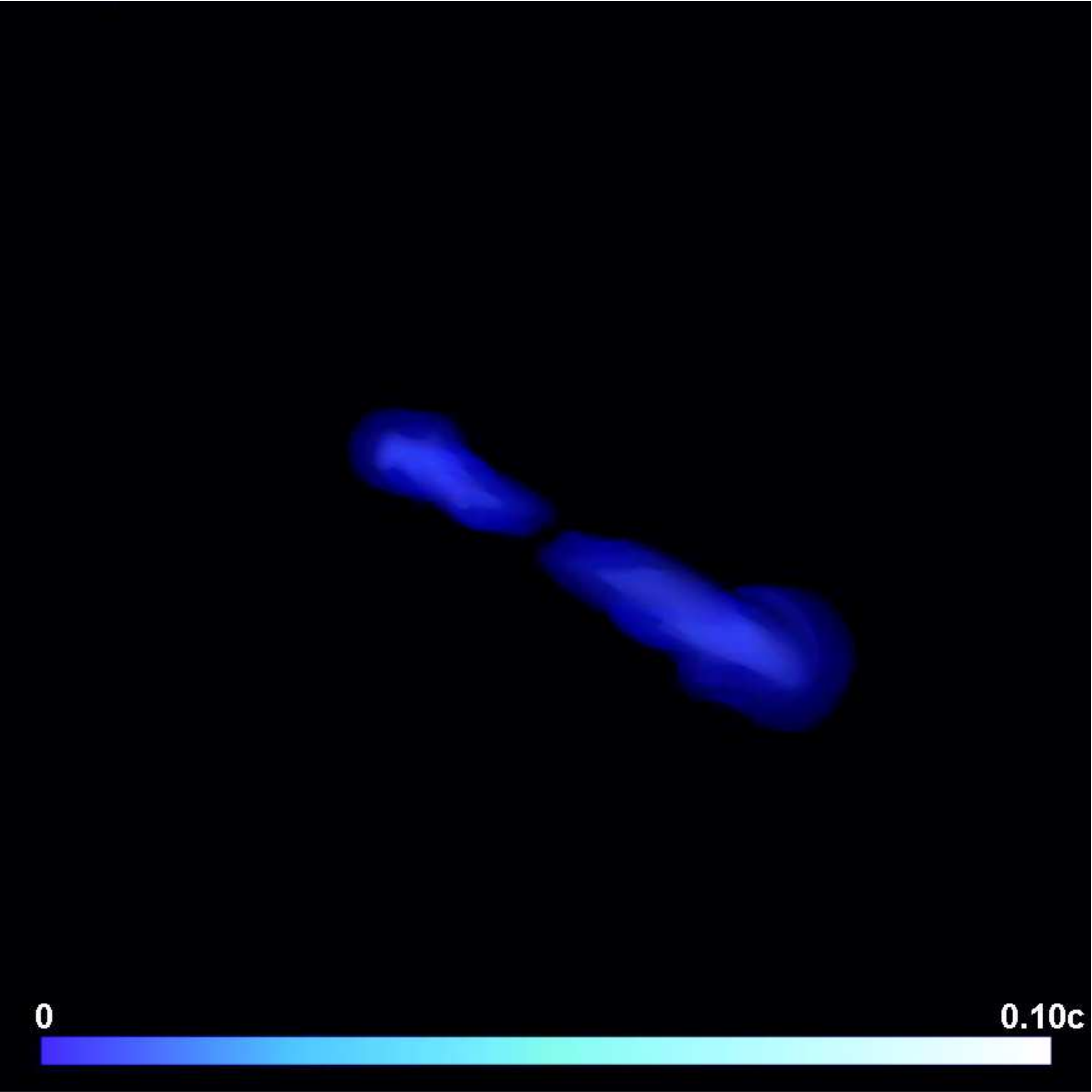}
\caption[Volume rendering of flow speed for all models]{Volume rendering of flow speed for the ST ({\it upper left}), I26 ({\it upper right}), I13 ({\it lower left}), and RE ({\it lower right}) models, after the jet disturbances have nearly reached the computational grid boundary.  The contrast in the RE image has been enhanced to increase visibility.  Animations of these quantities as seen from several different angles will be available through the electronic version of this paper.}
\label{fig:STspvr}
\end{center}
\end{figure*}

\subsection{Bi-directed Jets}\label{sec:jets}
Jet inflow is introduced entirely along the $\pm x$-axis without any initial transverse component.
The previously mentioned internal jet boundary consists of a cylindrical region originating from $x = 0$ and extending $6$ kpc along $\pm x$.
The resulting jets have uniform cores of radius $r_j = 3$ kpc, surrounded by a concentric transition annulus that smoothly connects to the ambient conditions at a radius of $7$ kpc.
The nominal core density contrast is $\eta = \rho_j/\rho_a = 0.01$, and the jets initially enter in pressure equilibrium with their surroundings.
The jet core magnetic field in this set of models is completely toroidal, of the form $B_{\phi} = B_0(r/r_j)$ within the jet core, but that decreases quadratically to zero outside of the core region.
This jet field was chosen to be strictly toroidal so that it could be initially divorced from the ambient field.
The fiducial ratio of thermal pressure to magnetic pressure in the jet (measured at the boundary of the core) is $\beta \sim 100$, and the associated (maximum) inflowing jet field strength is $B_0 = 10~\mu G$.

For this set of models, the jet luminosity can be expressed in terms of the steady jet values of the density $\rho_j$, velocity $v_j$, and pressure $p_j$ as

\begin{equation}
L_j \approx 2.12 \times 10^{45} \left[1.90\left(\frac{\rho}{\rho_j}\right)\left(\frac{v}{v_j}\right)^3 + \left(\frac{p}{p_j}\right)\left(\frac{v}{v_j}\right)\right] {\rm~erg~s}^{-1},
\label{eq:powerset2}
\end{equation}

\noindent ignoring the relatively minor contributions of magnetic and gravitational energy to the jet luminosity.
Since it is the velocity (and not the luminosity) that goes smoothly to zero in the transition annulus, the size of the annulus enters into the kinetic and thermal energy flux terms differently, resulting in the prefactor of 1.9 (see the Appendix for details).
This analytic expression is more than $99\%$ accurate over the lifetime of a steady jet.
The first jet model features two steady Mach 3 jets (labeled 'ST' for 'steady') that correspond to a physical flow speed $v_j \sim 0.1c$ and jet luminosity $L_j = 6.1 \times 10^{45}~{\rm erg~s}^{-1}$ per jet.
The other runs in this series of simulations feature jets with variable activity.
In all cases, the maximum injected jet speed and Mach number match those of the ST model, and it is worth noting explicitly that these assumed jet luminosities and speeds (corresponding to a Lorentz factor of 1.004) are consistent with a non-relativistic treatment.
In two runs, the jets intermittently switch on or off at regular intervals.  
In these two cases, the jets switch states every 26 Myr (labeled 'I26') and 13 Myr (labeled 'I13'), respectively.
This switch is accomplished by ramping the density, pressure, and momentum density exponentially over a 3.27 Myr (1.64 Myr) timescale for the I26 (I13) run.
The exponential ramp changes the density and pressure to a volume-averaged value sampled from a spherical region immediately surrounding the jet, allowing for a smooth transition from active jets to a quiescent state.
In the fourth model (labeled 'RE' for relic), designed to mimic radio relic sources, the jets are completely shut down after 26 Myr and are not restarted.
Additionally, there is a fifth model (labeled 'AM' for 'ambient') in which the jets are never activated.
This final model allows us to separate the evolution of the ambient medium from jet-driven behaviors to correctly assess the role of the environment in these simulations.

Figure \ref{fig:STspvr} shows volume renderings of flow speed at late times for each of the four jet-driven models.
In each frame, the jets originate near the center of the image at an angle with respect to the viewer, with the approaching jet pointing toward the lower right and the receding toward the upper left.
All four of the jet-driven simulations run until wave or shock disturbances from the jets reach a grid boundary. 
This timescale varies from model to model depending on how the flow is driven, and the total simulation times are given in Table 1.

\begin{figure}
\includegraphics[type=pdf,ext=.pdf,read=.pdf,width=0.5\textwidth]{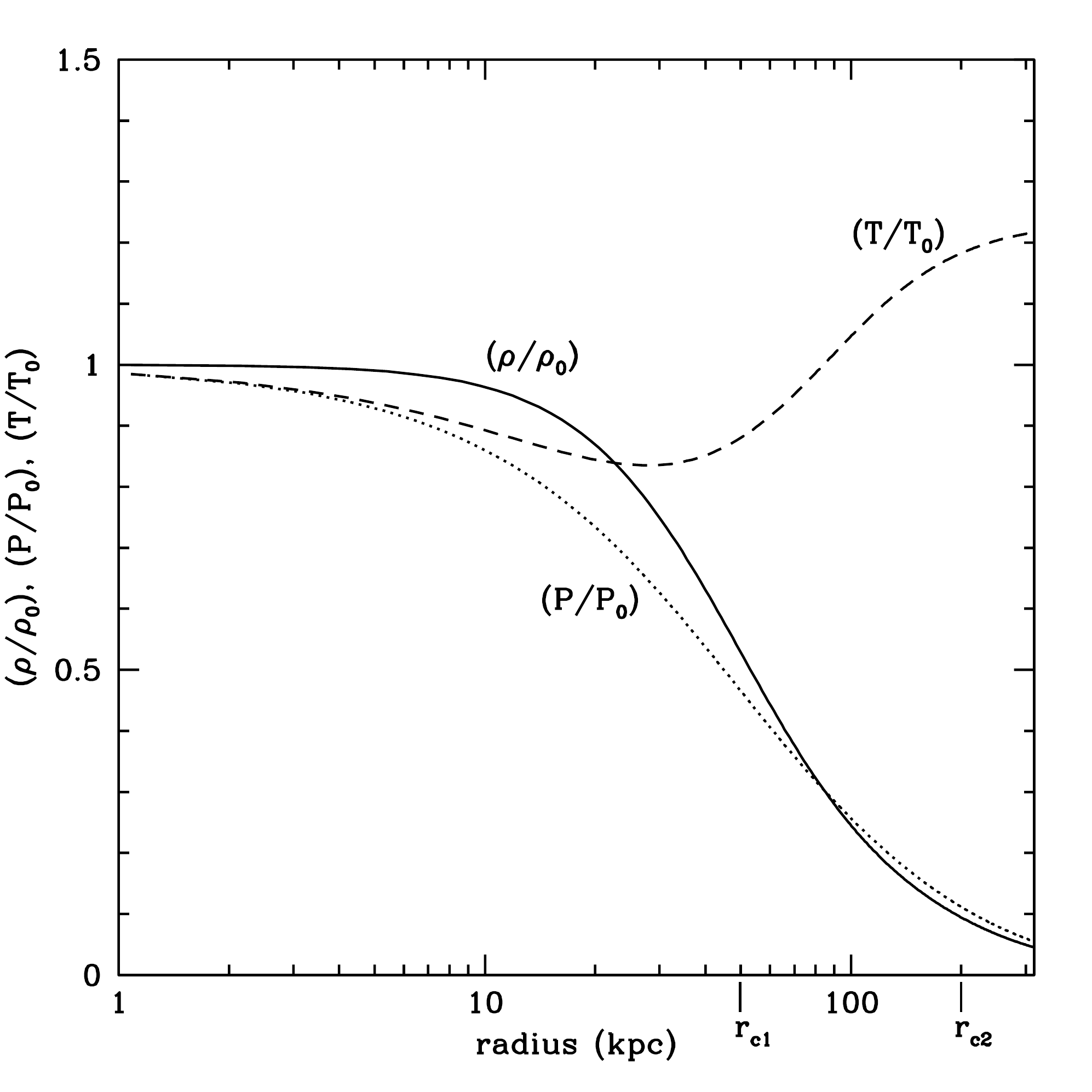}
\caption[ICM density, pressure, and temperature profiles]{ICM density, pressure, and temperature profiles.  Values of $\rho_0$, $P_0$, and $T_0$ are given in Table 1.  The labels $r_{c1}$ and $r_{c2}$ represent the locations of the ICM density core radii.}
\label{fig:ICMdenpret}
\end{figure}

\subsection{Galaxy Cluster Environments}\label{sec:ambient}
In all five models, we employ a single type of ambient medium designed to incorporate the essential properties of galaxy cluster environments.
The gas density base profile is a double-$\beta$ model (\eg \citealt{otamitsuda04}), similar to those typically used to observationally describe clusters, of the following form (see Figure \ref{fig:ICMdenpret}):

\begin{equation}
\rho_a(r) = \rho_0\left[\frac{f_1}{(1+(\frac{r}{r_{c1}})^2)^{\frac{3\beta}{2}}} + \frac{f_2}{(1+(\frac{r}{r_{c2}})^2)^{\frac{3\beta}{2}}}\right],
\label{eq:doublebetaden}
\end{equation}

\noindent where $\rho_0 = 8.33 \times 10^{-26}$ g cm$^{-3}$ is the density at $r = 0$.
This profile features two density ``cores'' weighted differently: $r_{c1} = 50$ kpc with a weight $f_1 = 0.9$ and $r_{c2} = 200$ kpc with a weight $f_2 = 0.1$.
These values were chosen to be consistent with the best-fit values found through a statistical analysis conducted by \citet{otamitsuda04} on 79 X-ray clusters.
For these models, $\beta = 0.7$, which generates an asymptotic profile only slightly steeper than $1/r^2$ when $r>>r_{c2}>r_{c1}$.
On top of this profile, we have superposed a \citet{kolmogorov41} spectrum of density fluctuations with a maximum amplitude of $\pm 0.10\rho_a(r)$ locally.
This amplitude of adiabatic fluctuations in density corresponds to the range of pressure fluctuations observed by \citet{schueckeretal04} in the Coma cluster. 
This initial spectrum of fluctuations is spatially tapered by a Gaussian envelope near the edges of the computational grid to impede the development of unphysical flows at the grid edges.

\begin{figure}
\includegraphics[type=pdf,ext=.pdf,read=.pdf,width=0.48\textwidth]{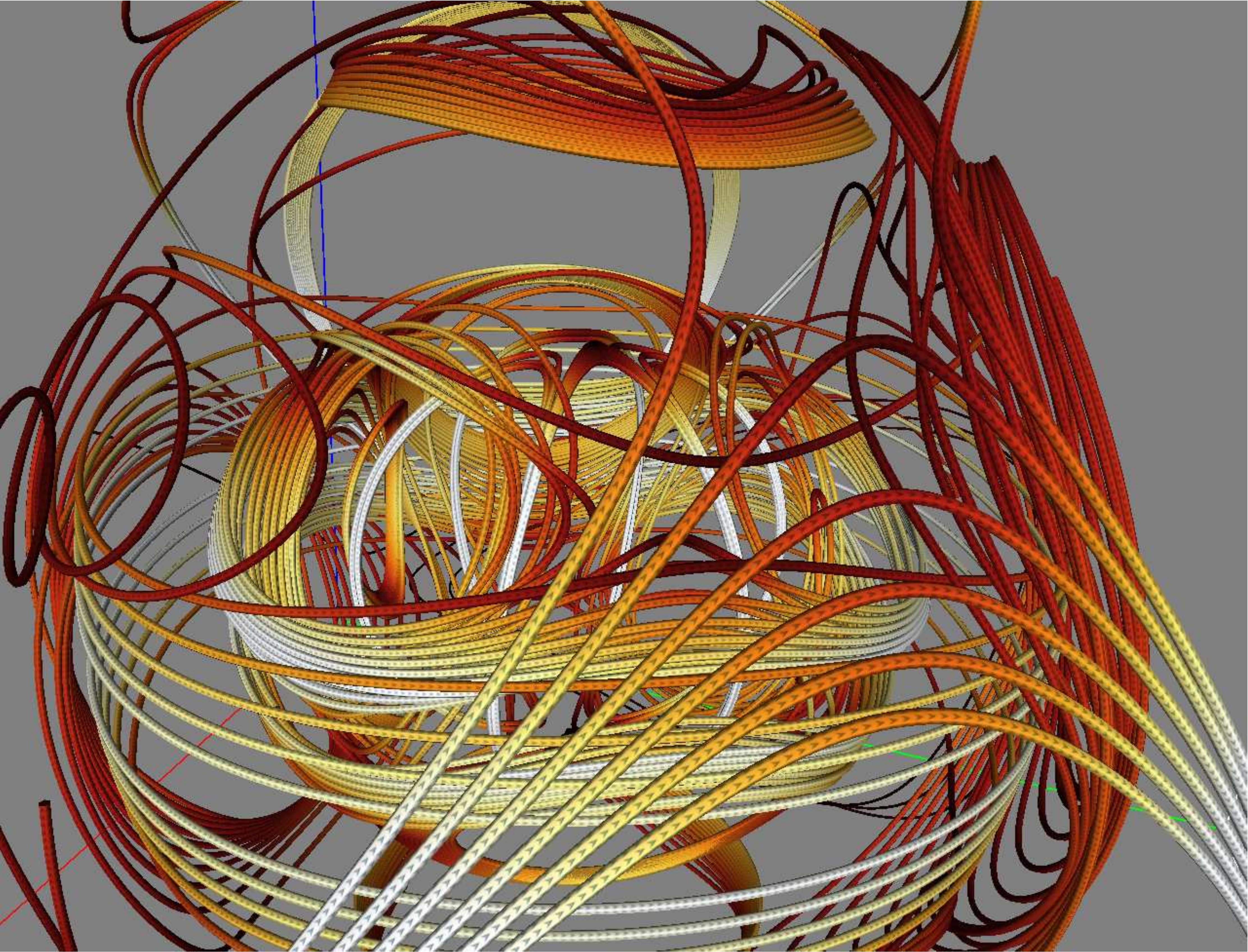}
\caption[Initial tangled ICM magnetic field lines]{Initial tangled ICM magnetic field lines.  Arrows on field lines indicate direction while intensity reflects field strength (light: stronger, dark: weaker).  Jets emerge from the region near the center of the image.}
\label{fig:fv_fieldt0}
\end{figure}

For these simulations, gravity is derived from an NFW profile \citep{nfw97}, intended to model the underlying cluster dark matter distribution:

\begin{equation}
\rho_{\rm dm} = \frac{\rho_s}{(\frac{r}{r_{\rm dm}})(1+\frac{r}{r_{\rm dm}})^2},
\label{eq:dmden}
\end{equation}

\noindent which generates a gravitational acceleration:

\begin{equation}
g(r) = -\frac{4 \pi Gr_{\rm dm}^3 \rho_s}{r^2}\left[ln\left(1+\frac{r}{r_{\rm dm}}\right)-\frac{r}{r+r_{\rm dm}}\right]
\label{eq:dmgrav}
\end{equation}

\noindent where $r_{\rm dm}=400$ kpc is taken to be the characteristic dark matter core radius.  
Another free parameter is the dark matter density, $\rho_s$.
Rather than deriving this from estimated cluster masses, we instead selected the value that would produce a reasonable temperature distribution (see below).
In this case, the temperature profile selected corresponds to $\rho_s \approx 4.3 \times 10^{-26}$ g cm$^{-3}$.
For a cluster with a virial radius $r_v = 2$ Mpc, this leads to a virial mass $M_v = 5 \times 10^{14} M_{\sun}$, which is within a factor of a few of the mass of the Perseus cluster ({\it e.g.} \citealt{ettorietal98}).

The pressure profile is determined by the requirement that the base atmosphere ({\it i.e.}, ignoring the density fluctuations) be in hydrostatic equilibrium.
The gravitational acceleration and gas density profile are sufficiently complex that the pressure profile is derived numerically and is shown in Figure \ref{fig:ICMdenpret}.
The core value of the pressure is $P_0 = 4.0 \times 10^{-10}~{\rm dyne~cm}^{-2}$.
The pressure and density profiles can be combined to give a temperature profile (Figure \ref{fig:ICMdenpret}), which is used as mentioned to provide the normalization of gravity.
The temperature profile was selected to resemble typical clusters (\eg \citealt{vikhlininetal06}).
As Figure \ref{fig:ICMdenpret} shows, the atmosphere has a core temperature $T_c \approx 2.5$ keV (assuming a mean molecular weight $\mu = 0.5$), with a total temperature range of $\sim \pm 0.2~T_c$ over the computational grid.
The sound speed, with a profile identical to that of the temperature, features a core value $c_{0} = 895 {\rm~km~s^{-1}}$.

The ambient magnetic field (see Figure \ref{fig:fv_fieldt0}) is designed to be a tangled, yet analytically specifiable, structure of the following form:

\begin{equation}
\vec{B}=B_\theta \hat{\theta} + B_\phi \hat{\phi},
\label{eq:set2field1}
\end{equation}

\noindent where the two components are

\begin{equation}
B_\theta = \frac{F_1(r) \cdot m}{r} \sin \theta \cos (m\phi)
\label{eq:set2field2}
\end{equation}

\noindent and

\begin{equation}
B_\phi = \frac{F_2(r) \cdot n}{r} \sin (n\theta) - \frac{F_1(r)}{r} \sin (m\phi) \sin(2\theta).
\label{eq:set2field3}
\end{equation}

The constants $m$ and $n$ are chosen to allow the field to vary in $\phi$ and $\theta$.
In our simulations $m = n = 3$, to allow for several full field reversals over either angle.
$F_1(r)$ and $F_2(r)$ are, in general, arbitrary functions of $r$ that we use to approximate a constant $\beta$ atmosphere with fluctuations.
In practice, we choose $F_1(r) = r B_{a}(r) [1 + sin(a_{kr1} r)]$ and $F_2(r) = r B_{a}(r) [1 + sin(a_{kr2} r)]$ to provide field variation with radius.
The constants $a_{kr1} = 2\pi/100~{\rm kpc}^{-1}$ and $a_{kr2} = 2\pi/58~{\rm kpc}^{-1}$ are used to provide fields that vary on spatial scales comparable to or slightly larger than the width of the jet.
These scales should be interpreted as the maximum length scales over which the field varies in our clusters rather than the scales on which magnetic power is concentrated.
The strongest fields in our clusters vary over a few tens of kiloparsecs, closer to the values inferred for observed clusters in \citet{vogtensslin05}, for example.
$B_{a}(r)$ is simply the ambient field strength as derived from the pressure profile and the requirement that $\beta \approx 100$ over a radial average, as shown in Figure \ref{fig:alphaplot}.

\section{Discussion}\label{sec:disc}
\subsection{Energy Flows}\label{sec:energy}

\begin{figure}
\includegraphics[type=pdf,ext=.pdf,read=.pdf,width=0.5\textwidth]{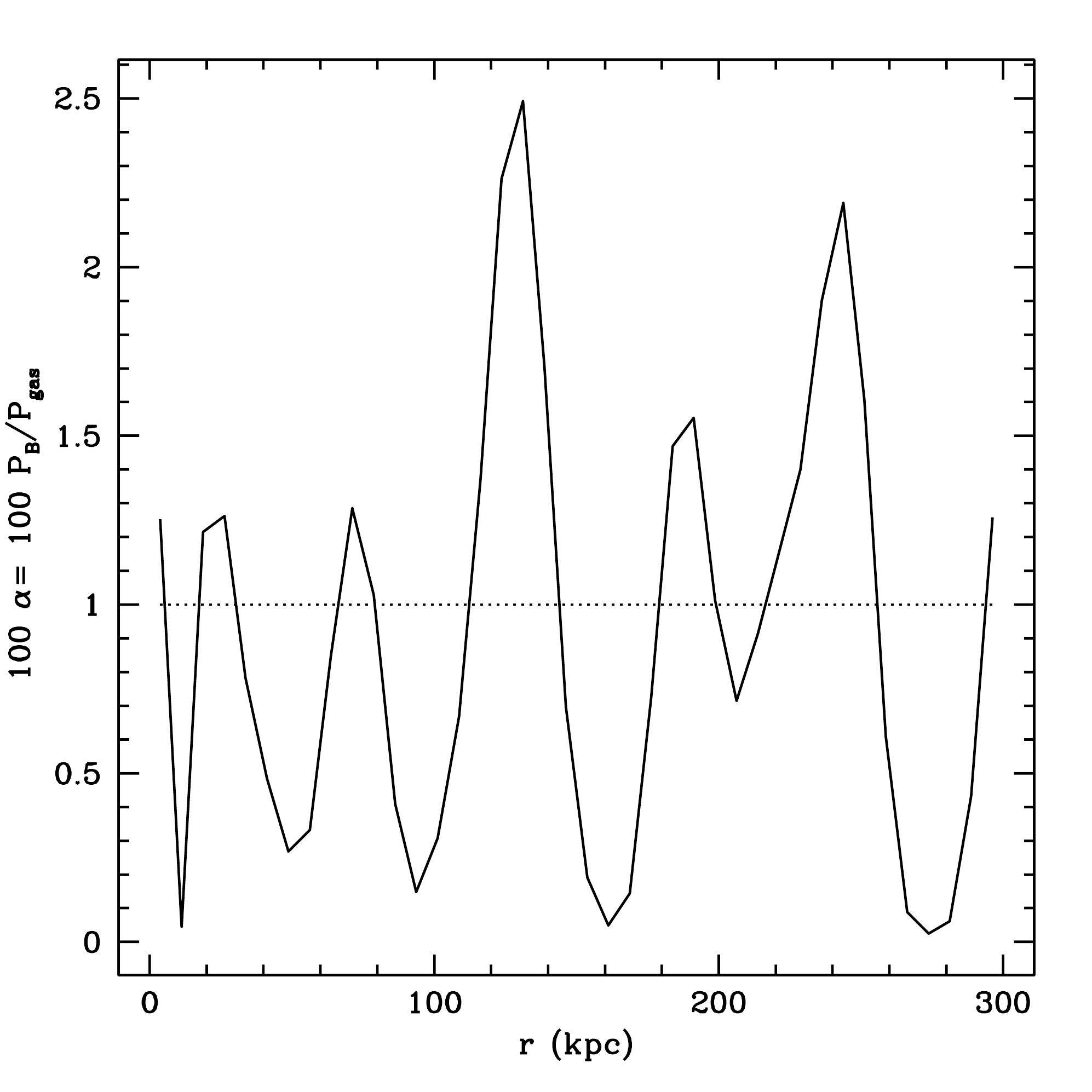}
\caption[ICM magnetic field variation]{Average values of $100 \alpha \equiv 100 P_B/P_{gas}$ (solid line), plotted as a function of radius from the cluster center.  The dotted line, shown for comparison, illustrates $100 \alpha = 1$, corresponding to $\beta \approx 100$.}
\label{fig:alphaplot}
\end{figure}

\begin{figure}
\includegraphics[type=pdf,ext=.pdf,read=.pdf,width=0.5\textwidth]{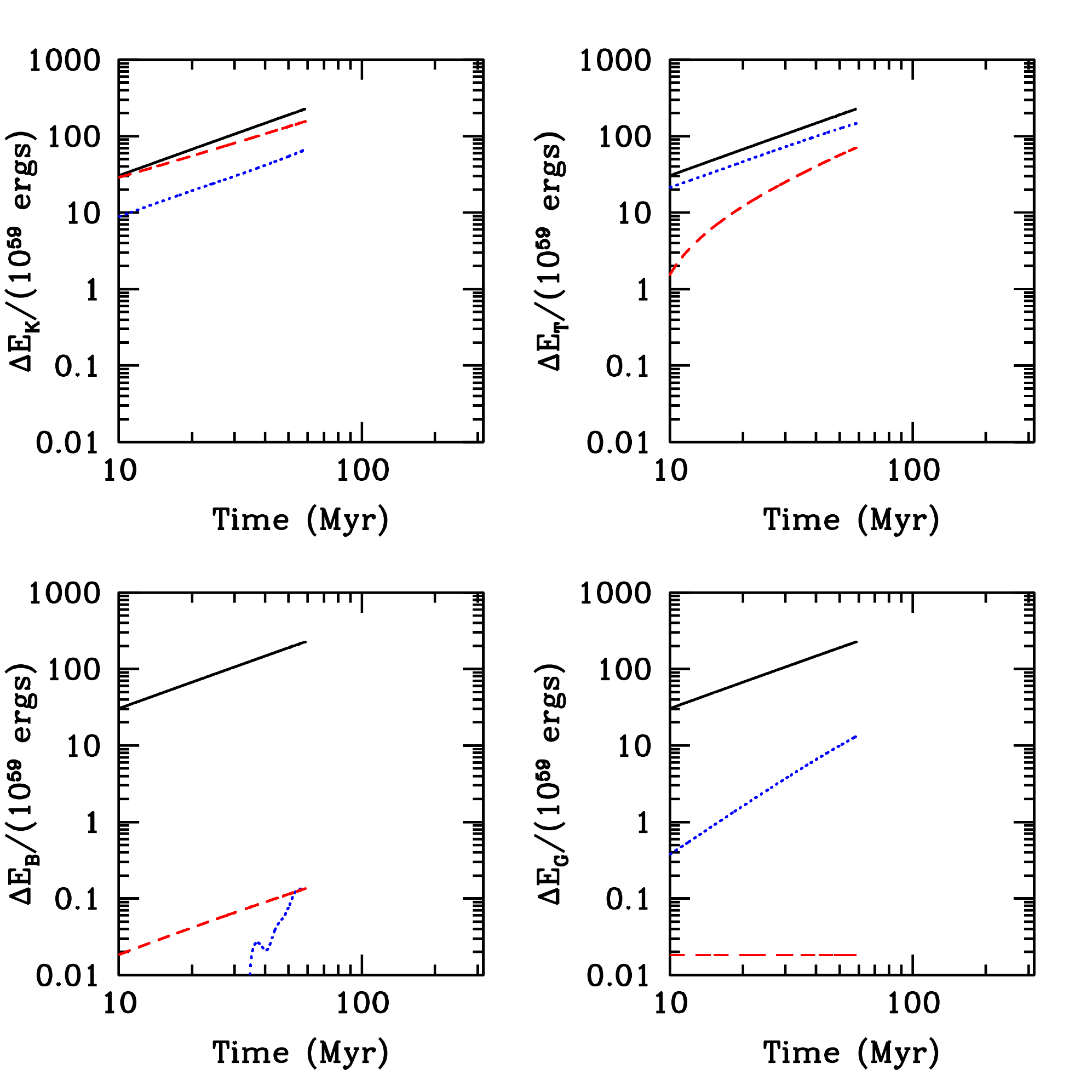}
\caption[Energy flow in the ST model]{Energy flow in the ST model.  {\it Upper left:} a comparison of the known added kinetic energy ({\em dashed line}) to the measured change (relative to the initial value) in kinetic energy on the grid ({\em dotted line}) as they vary in time.  The total (kinetic + thermal + magnetic+gravitational) inflow energy is shown ({\em solid line}) as a reference.  {\it Upper right:} same for the thermal energy.  {\it Lower left:} same for the magnetic energy.  {\it Lower right:} same for the gravitational energy.}
\label{fig:ST_eplot}
\end{figure}

We begin our discussion with an examination of energy flows in these systems.
By carefully tracking the jet power over time and comparing the integrated power to the measured change in energy on the grid for a given energy type, we can determine how energy is converted as the flow evolves.
Additionally, we can use the passive color tracer $C_j$ to indicate what fraction of this added energy enters the ICM.
We are particularly interested here in the amount of thermal energy transferred to the ICM since this energy is most likely to be of relevance to the quenching of cooling flows.
Again, we neglect radiative cooling; our primary goal is to understand energy transfer from the jets.

\begin{figure}
\includegraphics[type=pdf,ext=.pdf,read=.pdf,width=0.5\textwidth]{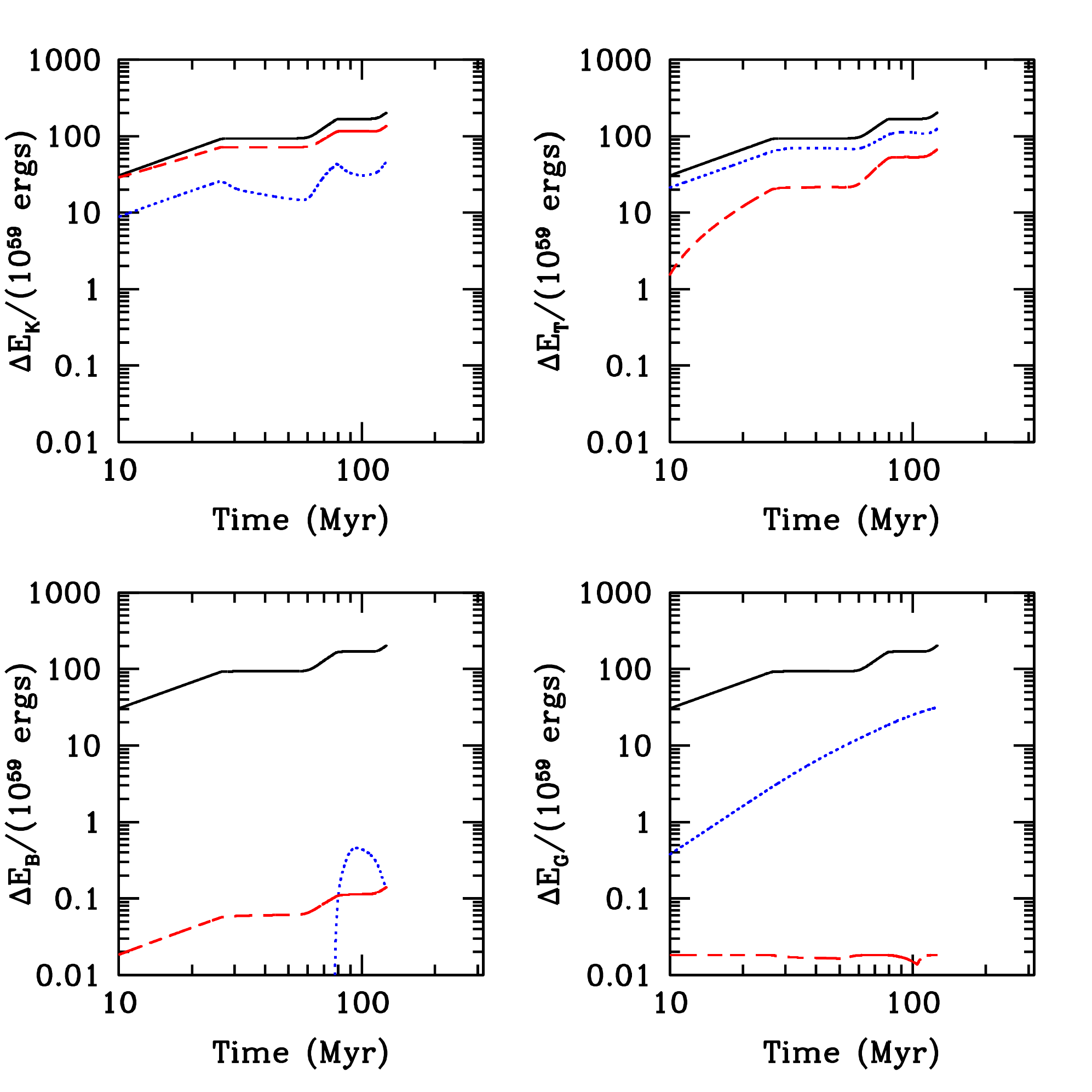}
\caption[Energy flow in the I26 model]{Same as Figure \ref{fig:ST_eplot}, but for the I26 model.}
\label{fig:I26_eplot}
\end{figure}

\begin{figure}
\includegraphics[type=pdf,ext=.pdf,read=.pdf,width=0.5\textwidth]{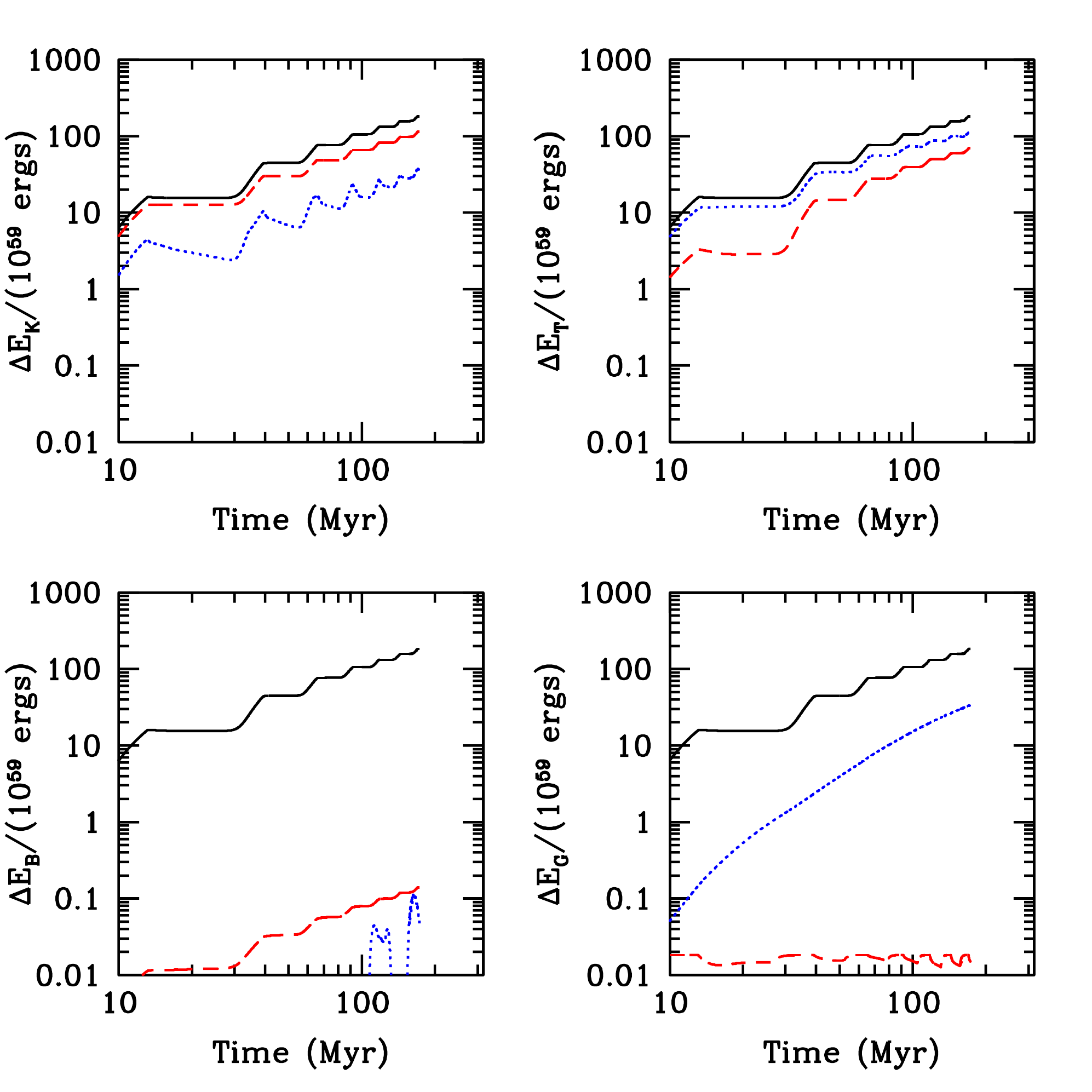}
\caption[Energy flow in the I13 model]{Same as Figure \ref{fig:ST_eplot}, but for the I13 model.}
\label{fig:I13_eplot}
\end{figure}

\begin{figure}
\includegraphics[type=pdf,ext=.pdf,read=.pdf,width=0.5\textwidth]{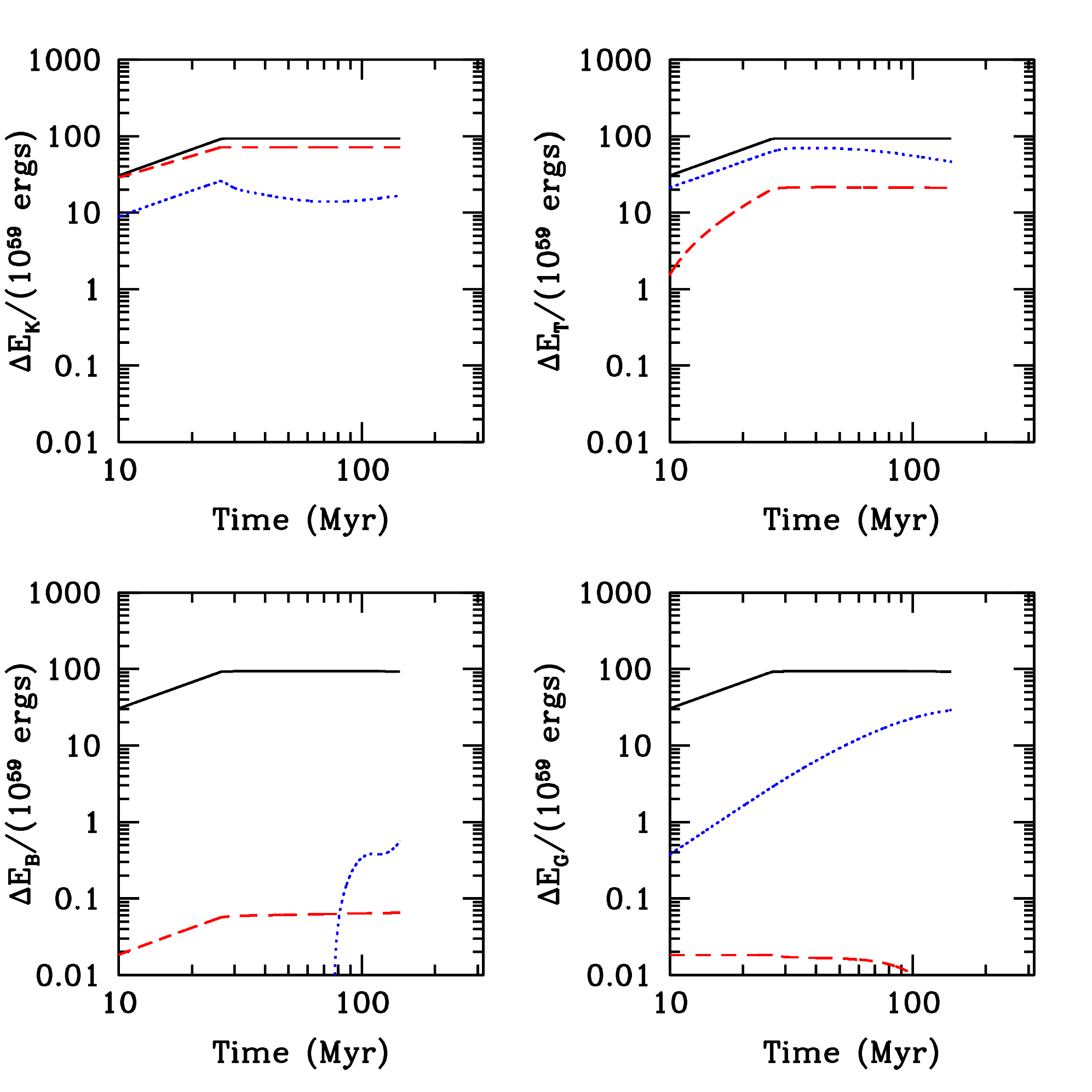}
\caption[Energy flow in the RE model]{Same as Figure \ref{fig:ST_eplot}, but for the RE model.}
\label{fig:RE_eplot}
\end{figure}

Figures \ref{fig:ST_eplot} - \ref{fig:RE_eplot} illustrate the energy flows for each of the four jet simulation models we use.
In each figure, four plots are shown, corresponding to the four types of energy present: kinetic, thermal, magnetic, and gravitational.
In each plot, the solid line represents the total amount of energy added to the grid by the jets, minus the energy measured to have exited the grid through the outer boundaries.
In these models, the latter quantity is very small since the systems would be in hydrostatic equilibrium were it not for the evolution of the added density perturbations and magnetic field fluctuations.
In practice, the amount of energy flux across the outer boundaries is further reduced by the fact that the density, pressure, and magnetic field are at their weakest in these regions.
The dashed line in each plot shows the integrated jet power for that particular energy type.
The dotted line shows the measured change in that energy type on the computational grid, excluding the jet inflow region that we treat as a boundary.

\begin{figure}
\includegraphics[type=pdf,ext=.pdf,read=.pdf,width=0.5\textwidth]{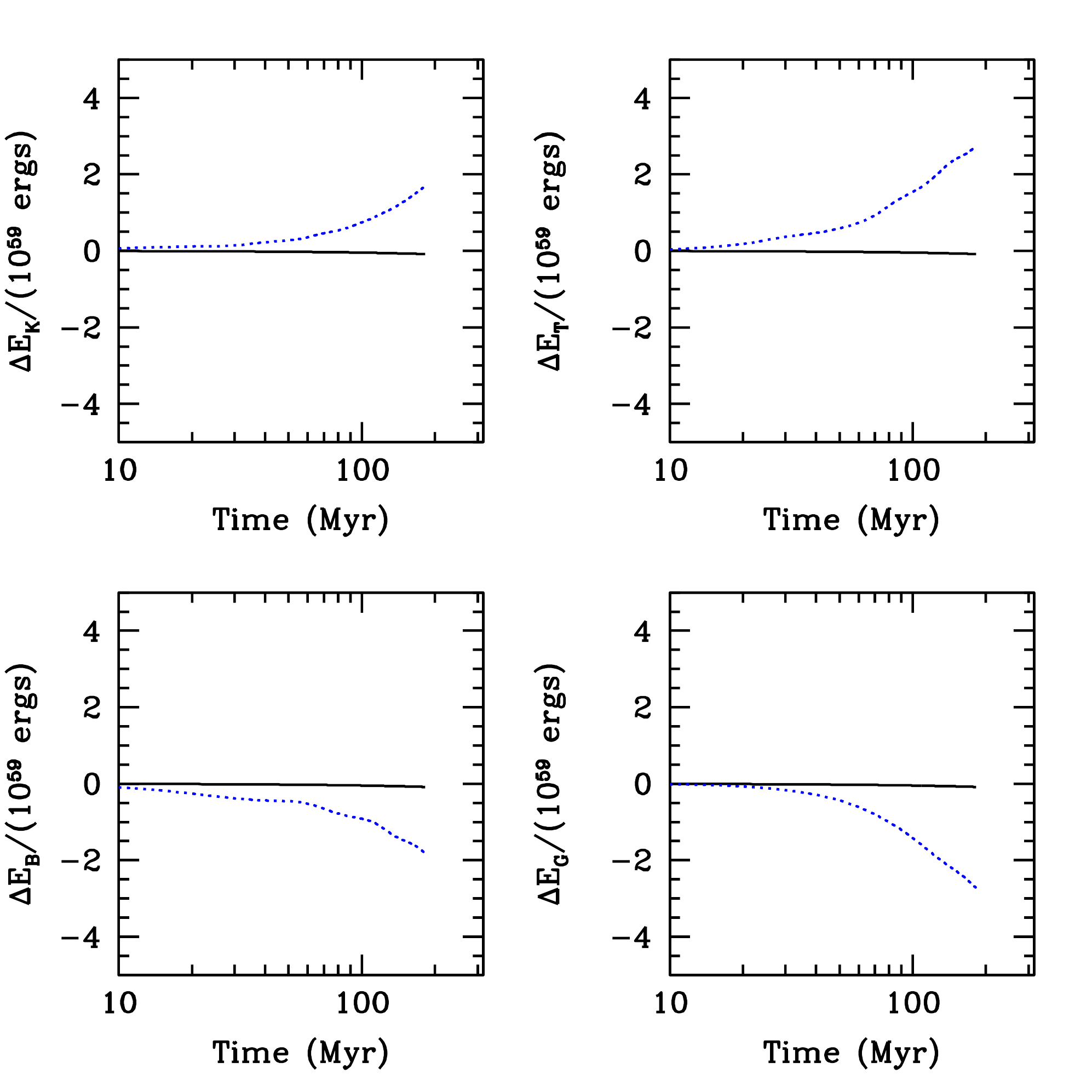}
\caption[Energy flow in the AM model]{Same as Figure \ref{fig:ST_eplot}, but for the AM model.  In this case, no jet inflow is shown since the jets are never activated.}
\label{fig:AM_eplot}
\end{figure}

Since the systems were not in perfect equilibrium initially, the ambient medium relaxes as the simulations progress, causing some exchange of energy independent of jet action.
Really, we are interested in energy flows resulting from the jet-ICM interactions rather than those established by the details of the ICM initialization, so this background relaxation is subtracted away from the measured change in energy ({\it dotted lines}) in each of Figures \ref{fig:ST_eplot} - \ref{fig:RE_eplot}.
To calibrate this background energy flow, we used the results of the AM model, which featured no jets. 
Figure \ref{fig:AM_eplot} shows the energy flow evolution for this model.
The solid line in each plot in Figure \ref{fig:AM_eplot} illustrates the accuracy with which we account for energy changes on the grid (the expectation value is zero), while the dotted lines show the change in each energy type as usual.
No dashed lines are included since no jet inflow was introduced.
As expected, the AM model relaxes by gaining kinetic and thermal energy at the expense of gravitational and magnetic energy.
The amounts of energy that change form in the AM model are only approximately $1\%$ of the total added jet energy in the ST model, but this amount is worth tracking since it exceeds the total added jet magnetic energy in that model, for example.

Focusing now on a discussion of Figures \ref{fig:ST_eplot} - \ref{fig:RE_eplot}, we point out some general features of energy flow in these simulations.
Unsurprisingly, the inflowing jet energy in all cases is dominated by the kinetic energy component (illustrated by the dashed line being just below the solid line in the upper-left plot in Figure \ref{fig:ST_eplot}, for example).
Most of the remainder of the energy added by the jets is in the form of thermal energy, with relatively minor magnetic and gravitational components.
The gravitational component ({\it dashed line in lower-right plot}), in particular, is roughly constant for active jets, representing the introduction of jet material in the gravitational potential upon jet initialization.
We note that the thermal component of the inflowing jet power ({\it dashed line in upper-right plot}) asymptotically reaches a somewhat larger fraction of the total jet luminosity for all active jets as time goes on.
This is caused by the sides of the jet cylinder allowing material back across that boundary, and is thus an artifact of the way in which we launch our jets.
The calculated jet power is reduced by this amount, however, so that the depictions of jet power are an accurate representation of the net inflow from the jet region.

\begin{figure}
\includegraphics[type=pdf,ext=.pdf,read=.pdf,width=0.5\textwidth]{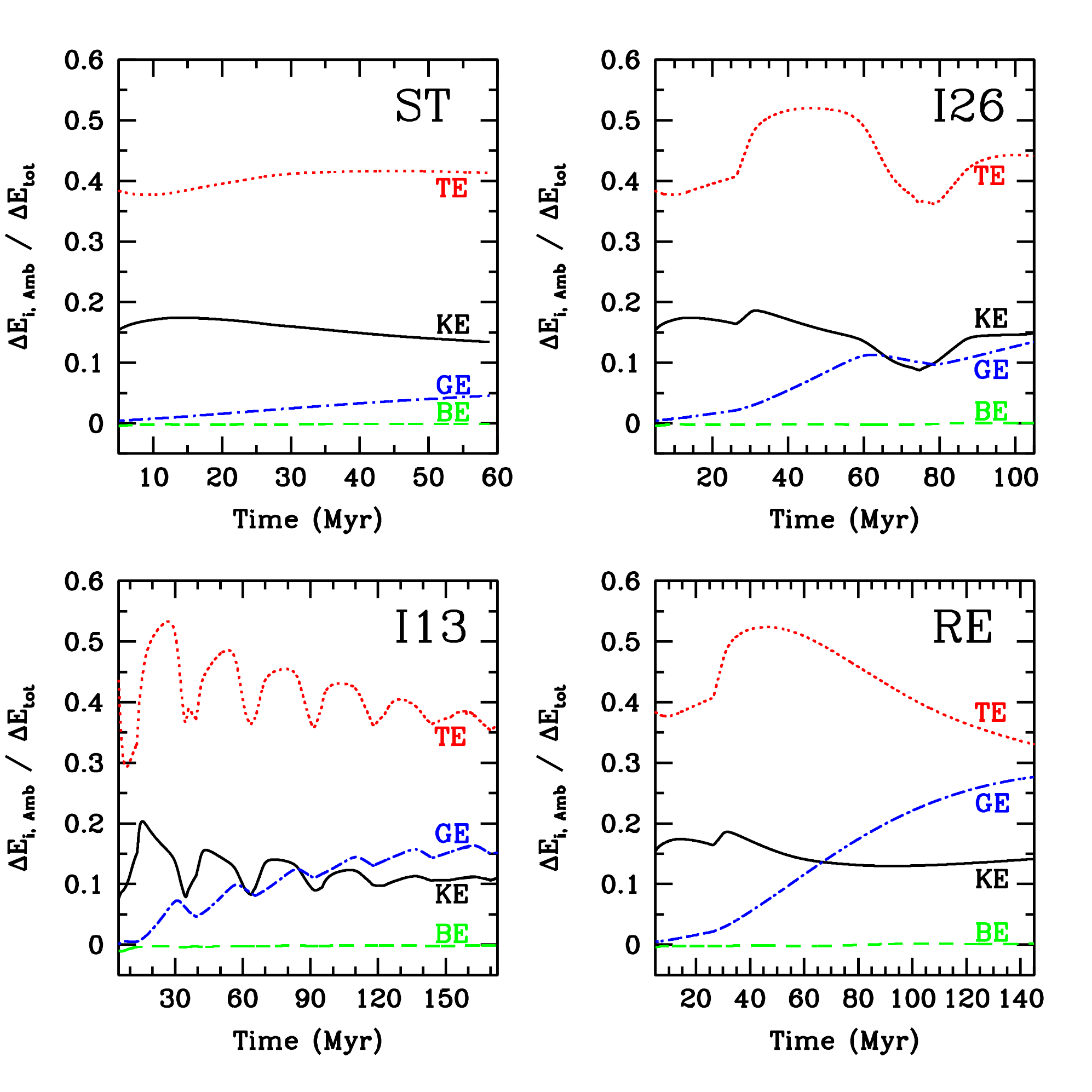}
\caption[Fractional energy deposited by jets into the ICM for each model]{Energy deposited by jets into the ICM over time, as a fraction of total energy added to the grid.  TE: thermal, KE: kinetic, GE: gravitational, BE: magnetic.  The model name is indicated in the upper-right portion of each plot.}
\label{fig:ST_efrac}
\end{figure}

There are also several global trends present in the measured changes on the grid ({\it dashed lines}) for the jet-driven models.
First, we note that in all models the measured thermal energy exceeds any of the other components.
Regardless of the details, all models generate substantially more thermal energy than is introduced through the jets.
Additionally, we note that all models feature appreciable increases in the gravitational energy.
Unlike our previous work \citep{oneilletal05}, these simulations include a realistic model of gravity in which the total gravitational energy on the grid is comparable to the total thermal energy.
Since little gravitational energy is introduced at the jet orifice, the increase in gravitational energy is almost entirely a result of energy exchange in these systems.
Finally, the measured magnetic energy evolution also features an interesting shape, suddenly increasing in each figure.
This is representative of the fact that the measured magnetic energy change on the grid is actually negative at early times and so is not shown on these plots; a point we will discuss shortly.

\begin{figure*}[t]
\begin{center}
\includegraphics[type=pdf,ext=.pdf,read=.pdf,width=0.9\textwidth]{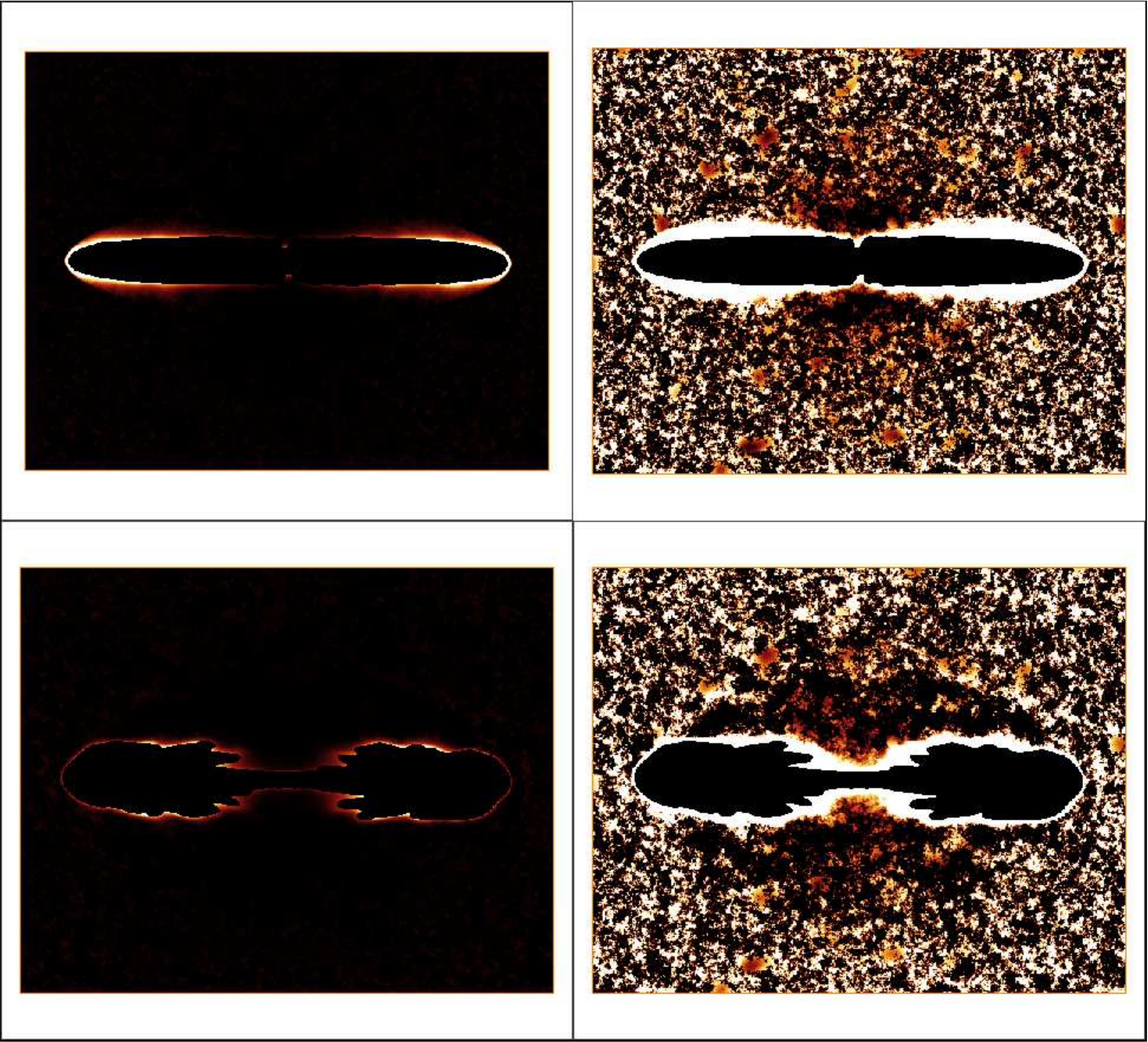}
\caption[Color-selected entropy density cross-sections for the ST and I13 models]{Color-selected entropy density ($T/\rho^{2/3}$) cross sections for the ST ({\it upper panels}) and I13 ({\it lower panels}) models at comparable jet lengths.  Entropy is only shown for material that is at least $90\%$ ICM by mass, and the initial entropy is subtracted off.  Higher intensities correspond to higher entropies, although the intensities in the right two panels are saturated to permit visualization of the influence of the bow shock.}
\label{fig:entropycomparisons}
\end{center}
\end{figure*}

Before we discuss the details of each model, we should first introduce another measure of energy flow in these systems.
By incorporating the passive color tracer, we can determine what fraction of the measured change in energy on the grid takes place in the ICM.
This is accomplished by measuring $(1-C_j)\Delta E$, where $C_j=1$ for jet material and $C_j=0$ for the ICM.
Figure \ref{fig:ST_efrac} shows, for the four jet models, this fraction for each different type of energy.
Again, the background change in energy due to the relaxation of the ICM is removed so that the included changes are ultimately the result of jet-ICM interactions.
The most significant trend to note is that all models are efficient at transferring energy from jets to the ICM.
As we found in \citet{oneilletal05}, efficient energy transfer seems to be a general characteristic of these systems, and all jet models have transmitted $\sim 60\%$ or more of their energy to the ICM by the end of the simulations.
Significantly, thermal energy always represents the largest fraction of energy added to the ICM for all models.
Unsurprisingly, given the large $\beta$ in the jet inflow, the magnetic energy always represents the smallest fraction of energy added.
The exact distributions of energy vary according to model, however, so we will address them individually.

\begin{figure}
\includegraphics[type=pdf,ext=.pdf,read=.pdf,width=0.5\textwidth]{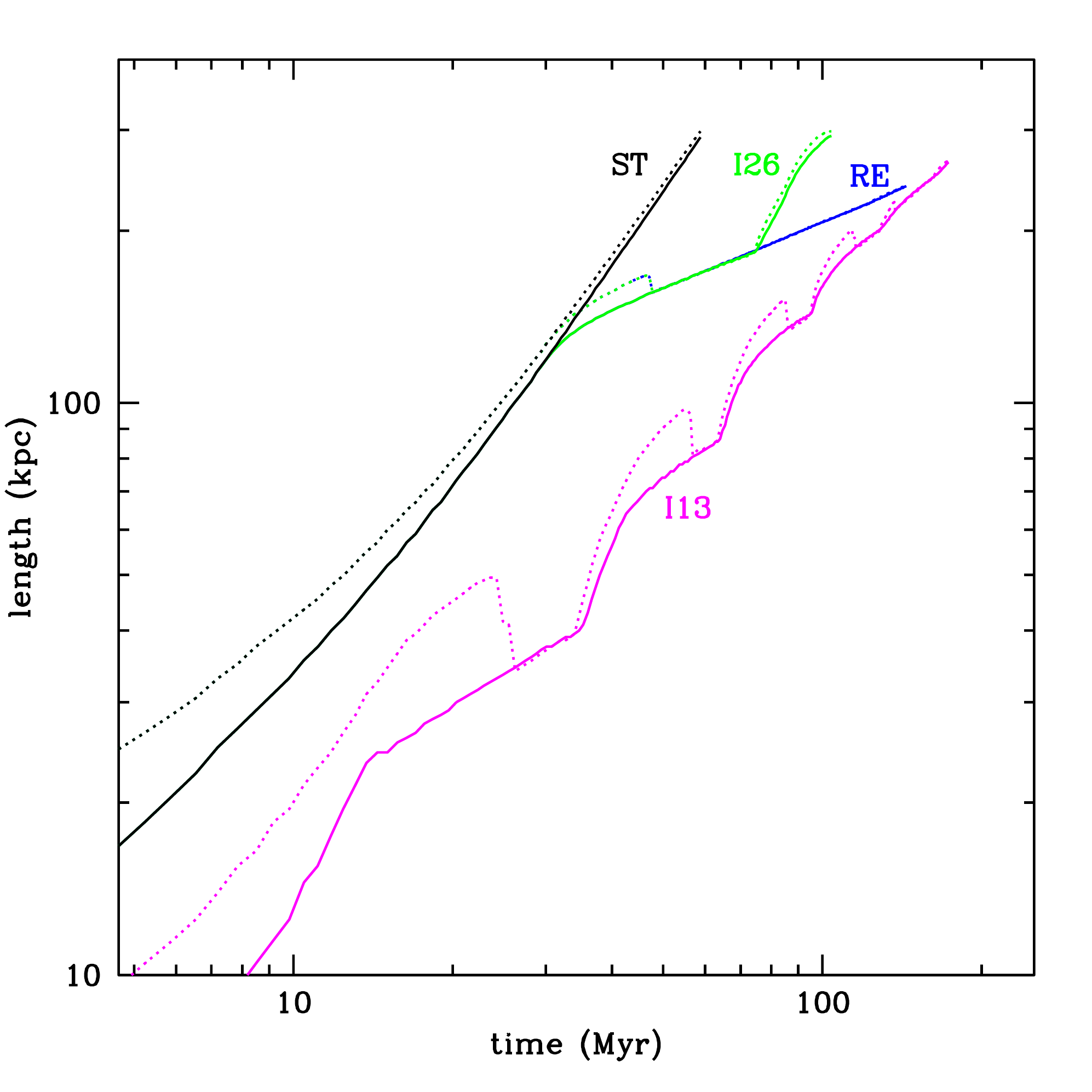}
\caption[The length evolution of our model jets]{Length evolution of our model jets.  Solid lines show the maximum extent in the $x$-direction as measured by the passive variable.  Dotted lines indicate the position of the bow shock as a function of time.  Values are averaged over the two jets in each simulation.}
\label{fig:doublejetlengths}
\end{figure}

Focusing for the moment on steady jets, represented by the ST model, there are several interesting features to note.
Similar to what we found in \citet{oneilletal05}, the active jets efficiently convert kinetic to thermal energy.
The measured thermal energy is asymptotically a factor of $\sim 2$ larger than thermal energy added by the jet, which is similar to the amount of kinetic energy lost from the jet inflow.
From the upper-left panel of Figure \ref{fig:ST_efrac}, we see that $\sim 40\%$ of the total energy added to the grid ends up as thermal energy in the ambient medium, suggesting that these steady jets efficiently heat their environments.
Although it is clear that this energy is being added to the ICM, we also seek to discover where in the ICM this energy appears.
As \citet{vernaleoreynolds06} and \citet{heinzetal06} demonstrated, the location of deposited energy is at least as significant as the amount of energy deposited for a system to offset cooling.
Figure \ref{fig:entropycomparisons} shows illustrations of the entropy density $s = T/\rho^{2/3}$ for the ST and I13 models.
Specifically, we apply the color tracer and show $\Delta s=s-s_0$ for $C_j < 0.10$, which removes the initial ambient entropy $s_0$ and focuses only on enhancements in the ICM. 
Looking at the upper-left panel, most of the visible $\Delta s$ forms a sheath around the jet.
This is caused by a mixture of jet-ICM mixing at the boundary (jets have low density, so at
a given pressure, high entropy) and ICM entropy increases resulting from shock heating.
In the saturated image ({\it upper right}), we can see the effects of the bow shock, although the shock discontinuity itself is not visible.
These images suggest that much of the irreversibly added energy measured in the ICM in the ST model probably is transferred and remains near the jets themselves rather than uniformly filling the shocked ICM region.

Figure \ref{fig:ST_eplot} also shows that the measured gravitational energy change represents an increasing fraction over time of the total added energy for steady jets, although it still comprises less than $10 \%$ of the total energy by the end of the simulation.
Since the jets are light with respect to the ambient medium, they contain a very small fraction of the total gravitational energy.
In fact, you would have to lift jet material $100$ times higher than core ICM material for the same energy gain.
Rather, this measured change is caused by the jets driving ICM material higher into the gravitational potential.
Figure \ref{fig:ST_efrac} also supports this interpretation since the fraction of gravitational energy added to the ambient medium ($\sim 5\%$) is comparable to the total fraction estimated from Figure \ref{fig:ST_eplot}.
The increase in gravitational energy can either be accomplished mechanically --- through lifting, dragging, or entrainment --- or through the ICM reacting to the added heat and local pressure.
In the ST model, dragging ({\it i.e.}, as in a wake) should not play a role since the jet is never disconnected from the source.
Since the passive variable appears to have a sharp boundary in this model (which you can see from the mass-fraction-selected entropy in Figure \ref{fig:entropycomparisons}), the ICM is also not significantly entrained.
Any lifted ICM material would have to be located between the jet/ICM contact discontinuity and the bow shock.
In the direction of jet propagation for an active jet, this is a very small region and so not likely to contribute much.
Since we know that the ICM is heated, this is the most likely source of the increase in gravitational energy for this model.

\begin{figure*}[t]
\begin{center}
\includegraphics[type=pdf,ext=.pdf,read=.pdf,width=0.45\textwidth]{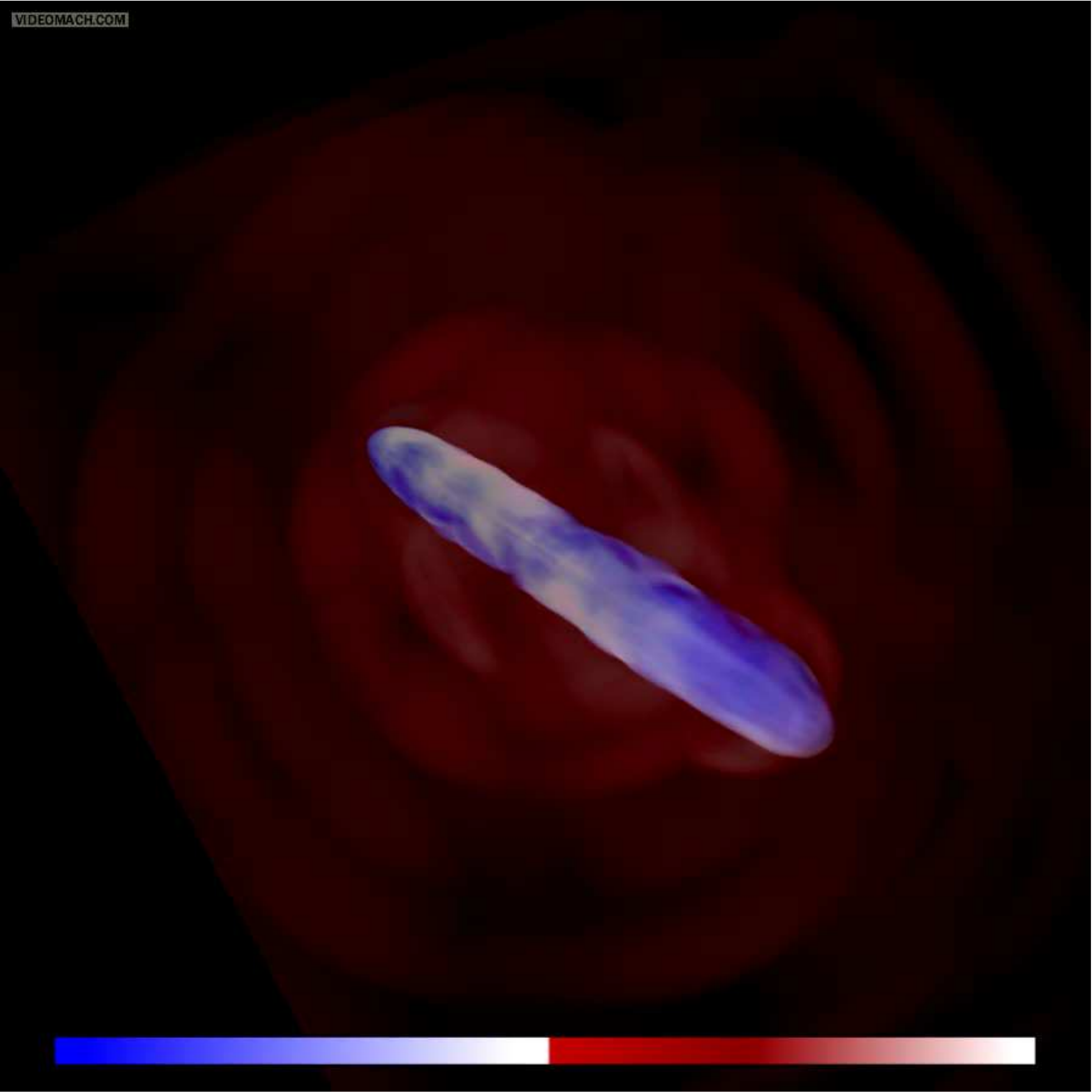}
\includegraphics[type=pdf,ext=.pdf,read=.pdf,width=0.45\textwidth]{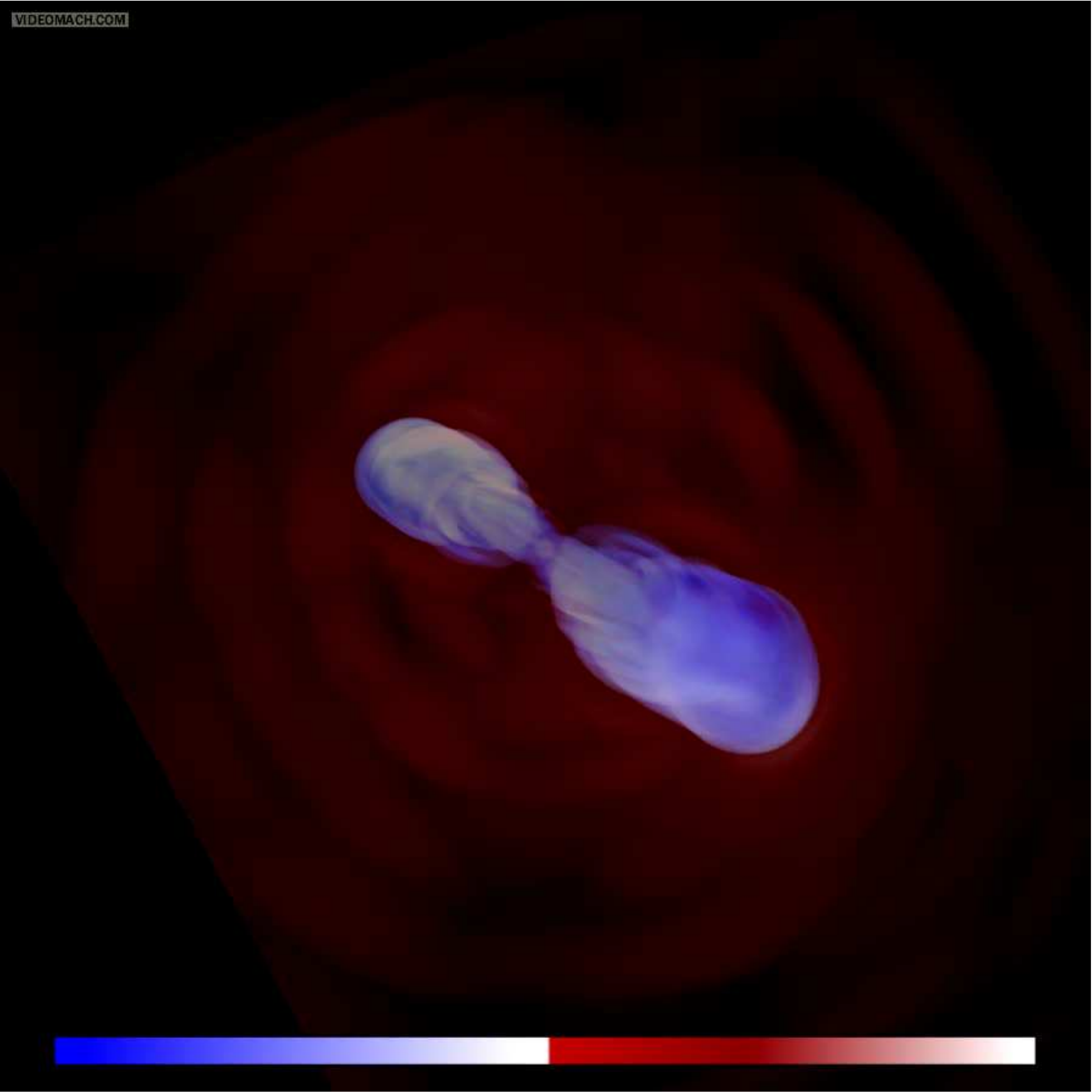}
\caption[Volume rendering of magnetic field strength for the ST and I13 models]{Volume rendering of the magnetic field strength for the ST ({\it left}) and I13 ({\it right}) models.  Lighter colors correspond to stronger fields.  Animations of these quantities as seen from several different angles will be available through the electronic version of this paper.}
\label{fig:ST_bm}
\end{center}
\end{figure*}

The evolution of the magnetic energy in the ST model is interesting since magnetic energy is initially lost from the system and later increases.
In fact, this is a general characteristic of all four jet models.
Although the total magnetic energy budget is tiny with respect to the other forms of energy (making the accurate subtraction of the ICM more crucial), we still want to understand how magnetic fields evolve, particularly since these fields are an essential component of the synchrotron emission we observe.
There are two possible physical scenarios one can imagine to achieve a net decrease in magnetic energy after the ICM relaxation has been subtracted.
One is that the stronger central ICM fields get advected upward to where they can more easily expand, reducing their magnetic energy.
This seems implausible since the ST model does not advect much ambient material without compressing it.
Also, the observed effect is almost immediate, and so operates at early times. 
The second possibility is that the ambient fields reconnect with the jet fields and then relax.
Although reconnection as a process would not likely lead directly to a significant loss of magnetic energy, it has the potential to create new field geometries that could facilitate magnetic relaxation.
This is the more plausible of the two scenarios since the jet fields certainly reconnect with the surrounding ICM (see the discussion in Section 3.3).
Unfortunately, it is less obvious that this process would necessarily lead to a net reduction in magnetic energy since the stretching and twisting of reconnected fields could also amplify the fields (see, for example, Figure \ref{fig:relicreconnection}).
After this initial dip, however, it is clear that the overall magnetic energy in the system is increasing.
Some of this is compression of the ICM magnetic field by the jet/cocoon/bow shock, but Figure \ref{fig:ST_bm} suggests that much of it is simply Poynting flux introduced by the jet inflow.
The animations associated with Figure \ref{fig:ST_bm} further suggest that the magnetic energy in the cocoon is replenished by the jet since the field in the cocoon is reduced in magnitude during periods of jet quiescence, presumably as a result of adiabatic expansion.

\begin{figure}
\includegraphics[type=pdf,ext=.pdf,read=.pdf,width=0.5\textwidth]{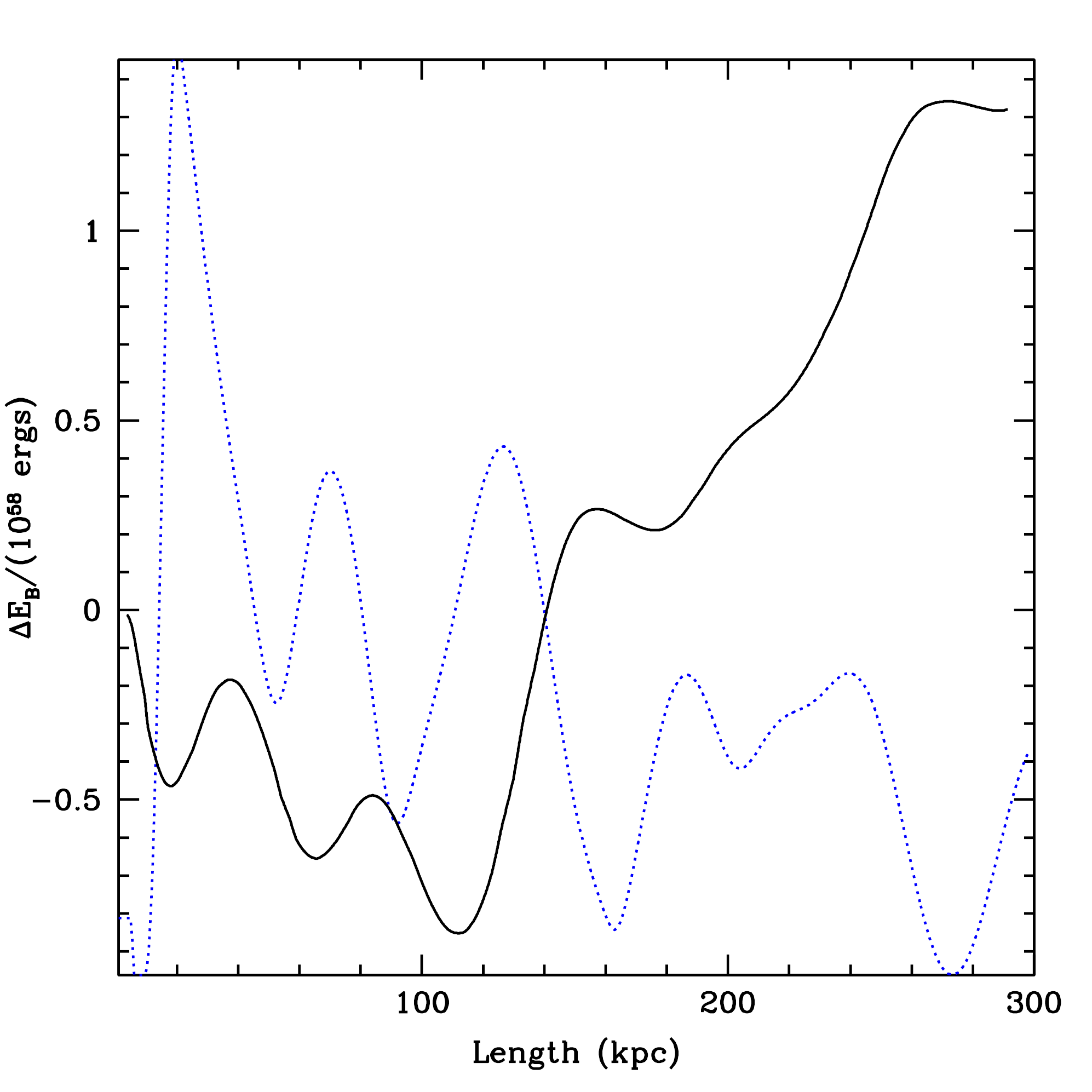}
\caption[Comparison of the magnetic energy evolution of the ST model to the ICM magnetic field structure]{A comparison of the magnetic energy evolution of the ST model to the ICM magnetic field structure.  The solid line, corresponding to the values on the $y$-axis, shows the change in magnetic energy on the grid in the ST model.  The dotted line (scaled for comparison) shows the magnitude of the ICM field in the direction of jet propagation.}
\label{fig:ST_beplot}
\end{figure}

Interestingly, there is a correlation between changes in magnetic energy and jet encounters with the structure of the ICM magnetic field.
As Figure \ref{fig:alphaplot} shows, the radial distribution of magnetic field is not uniform, but rather peaks in magnitude every $50~$kpc or so.
Using the measured jet lengths in Figure \ref{fig:doublejetlengths}, which we will discuss further in Section \ref{sec:dm}, we can translate the time evolution of magnetic energy to evolution as a function of jet length.
Figure \ref{fig:ST_beplot} shows the relationship between the change in magnetic energy and the ICM magnetic field in the direction of jet propagation.
The two quantities appear to be correlated --- or anti-correlated --- with a slight offset.
The causal reasons for this are interesting since it is not obvious that the structure of the globally weak ICM field should be detectable in the energy flow.
Animations associated with Figure \ref{fig:ST_bm} show the magnetic field evolution for the ST and I13 models.
In each case, it is clear that the ICM magnetic energy is enhanced each time the jet passes through a strong magnetic pressure gradient in the ICM.
These waves of enhancement lag slightly behind the jet intersection of a high-field region, suggesting that the solid line in Figure \ref{fig:ST_beplot} is correlated with, but lags slightly behind, the dotted line.
Since a substantial fraction of the magnetic energy on the grid is located in these regions, it makes sense that their perturbations would manifest themselves in the measured magnetic energy changes.

We turn now to a discussion of energetics in the two intermittent jet models, I26 and I13.
As Figure \ref{fig:doublejetlengths} and the animations associated with Figure \ref{fig:STspvr} illustrate, the appearances of the two intermittent jet models deviate from one another very early on.
This is due mostly to the fact that the I26 model manages to form a well-defined low-density jet channel before the first `off' cycle, whereas the I13 model must struggle at first to break through the dense ICM.
As Figures \ref{fig:I26_eplot} and \ref{fig:I13_eplot} show, however, the two models are energetically quite similar.
The kinetic and thermal energy influxes closely follow the level of jet activity.
As an example, the I26 jet is on for 26 Myr, corresponding to a monotonic increase in added kinetic and thermal energy, just like the ST model.
When it switches off from $t \sim 26-29~$Myr, the total added energy levels off while kinetic energy briefly continues to convert to thermal at the terminal shock.  
While the jet remains off from $t \sim 29-55~$Myr, the system loses kinetic energy and gains gravitational energy.
As the jet finishes restarting by $t \sim 60~$Myr, the cycle begins again, and similarly for the I13 model.
It is worth noting that the gravitational energy increases monotonically through the entire lifetime of the model. 
This is consistent both with the ICM reacting to the added heat (as in the ST model) and the dragging up of ICM material in the wakes of the unpowered, momentum-driven jet plumes.
This latter process almost certainly takes place in these models since we see the kinetic energy of the system decrease while the gravitational energy increases at $t \sim 29-55~$Myr.
The fractional energy plots (Figure \ref{fig:ST_efrac}) show that both models are very efficient at heating the ICM, although the fraction of energy in the ICM at a given time can vary by as much as $\sim 20\%$, depending on the level of jet activity.

One interesting point concerning the I13 model is that there are enough cycles of jet activity that we can attempt to discuss asymptotic behaviors.
Compared to the ST model, the I13 model has approximately the same total efficiency ($\sim 60\%$) in transferring energy to the ICM.
The fraction of kinetic and thermal energy entering the ICM is similar for the two runs, but the gravitational energy fraction is $\sim 10\%$ higher for the I13 model.
This is slightly misleading, however, since the {\it total amount} of gravitational energy introduced into the system at a given time is higher for the ST than the I13 model.
The fractional gravitational energy differences in Figure \ref{fig:ST_efrac} are thus more indicative of the difference in the total energy available to each model.
Additionally, we can again examine the entropy to see where the intermittent jets characteristically deposit their energy.
The bottom two panels in Figure \ref{fig:entropycomparisons} show the color-selected $\Delta s$ for the I13 model.
Although the added ICM entropy still forms a sheath around the jet cocoon, there is now a much more substantial amount of material nearer to the cluster core.
Likewise, as we will see in Section \ref{sec:dm}, the intermittent cocoons are characteristically fatter than those of the ST model.
This means that more high-entropy material is spread through a larger volume and nearer to the cluster core, both of which are needed to offset core cooling.

We turn lastly to the RE model, which evolves identically to the ST model until the jet activity is shut off completely at $\sim 26~$Myr.
In Figure \ref{fig:RE_eplot}, we see the expected inflowing and total energy plateaus appearing at $\sim 26~$Myr.
As in the I26 and I13 models, the system begins to exchange kinetic energy for gravitational energy as the thermal energy is approximately maintained.
After $t \sim 60~$Myr, however, we see the total thermal energy start to decrease as the kinetic energy mildly increases.
This feature is also marginally present in the I26 model, but in that case the jet reactivates before it becomes very pronounced.
In the RE model, the slope of the added gravitational energy also begins to flatten.
This behavior is consistent with the transition from a system actively driven by jets to one in which a passive plume consisting of a mix of ambient and jet/cocoon material is gradually slowed by the drag force associated with its motion through the ICM (see Section \ref{sec:dm} for details). 
As the relic plume rises, it should maintain approximate pressure equilibrium with the surrounding ICM, expanding and losing thermal energy in the process.
Likewise, reversible heat added to the ICM can be transformed back to kinetic energy through adiabatic expansion.
Ultimately, the total gravitational energy should decrease when the system becomes dominated by buoyancy, but there is still sufficient momentum in the plume through the duration of our simulations that we have not yet reached that regime.
Figure \ref{fig:ST_efrac} shows that approximately $\sim 75\%$ of the jet energy has entered the ICM by the end of the RE simulation.
Despite being more efficient at delivering what energy it has, however, Figure \ref{fig:RE_eplot} shows that the RE model introduces less total energy into the ICM than any of the intermittent or steady models as a result of its lower integrated luminosity.

\subsection{Dynamics and Morphology}\label{sec:dm}

\begin{figure*}[t]
\begin{center}
\includegraphics[type=pdf,ext=.pdf,read=.pdf,width=0.45\textwidth]{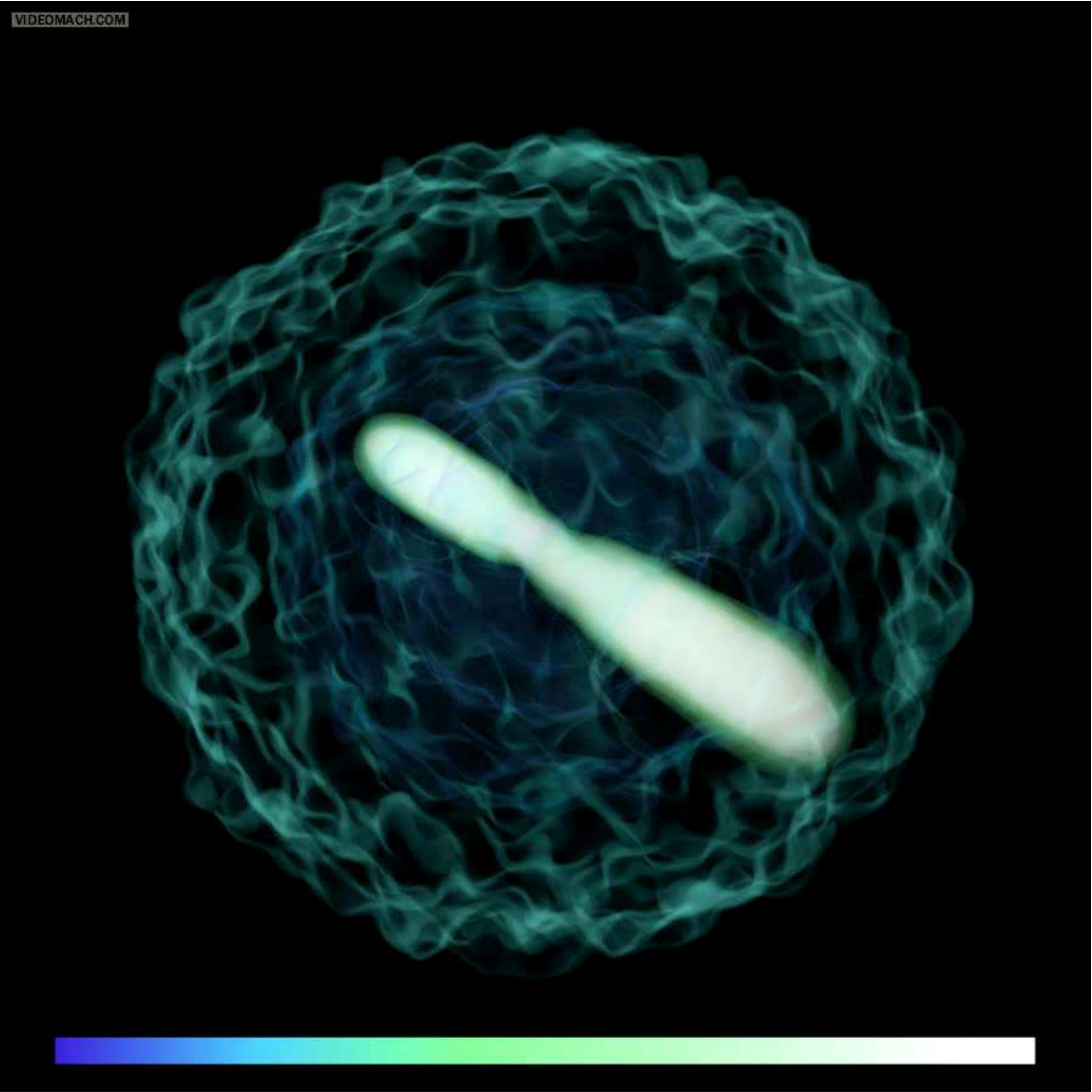}
\includegraphics[type=pdf,ext=.pdf,read=.pdf,width=0.45\textwidth]{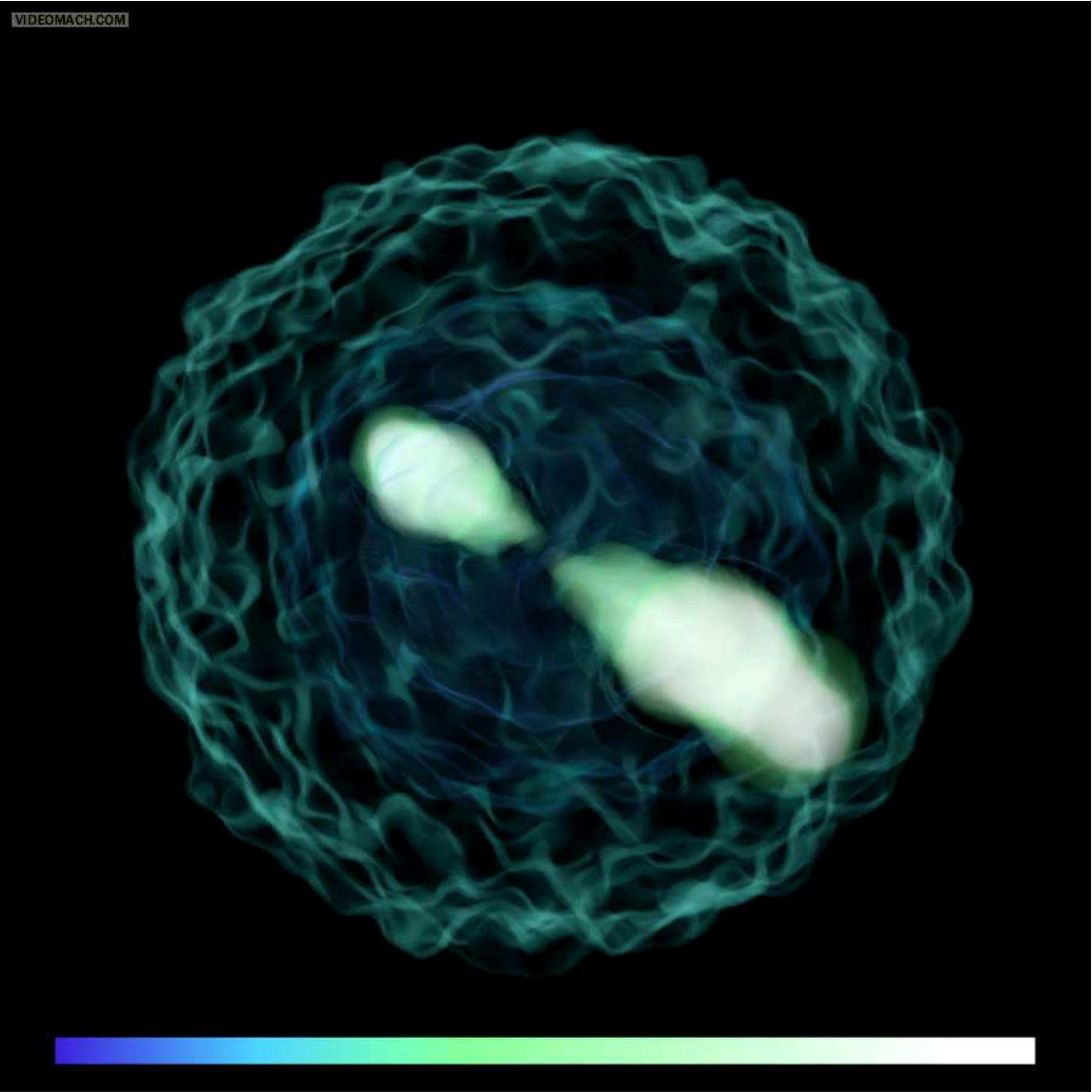}
\\[0.75em]
\includegraphics[type=pdf,ext=.pdf,read=.pdf,width=0.45\textwidth]{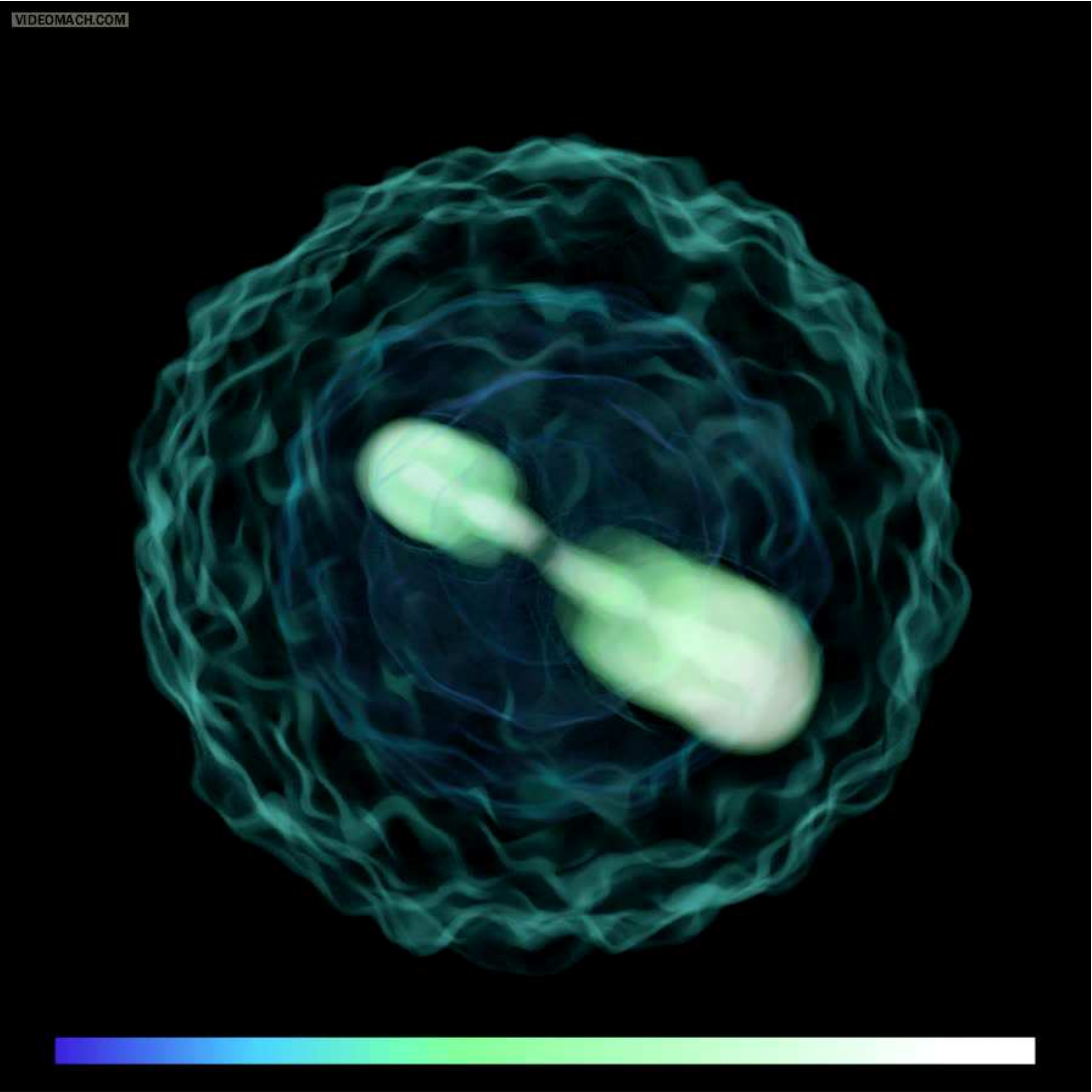}
\includegraphics[type=pdf,ext=.pdf,read=.pdf,width=0.45\textwidth]{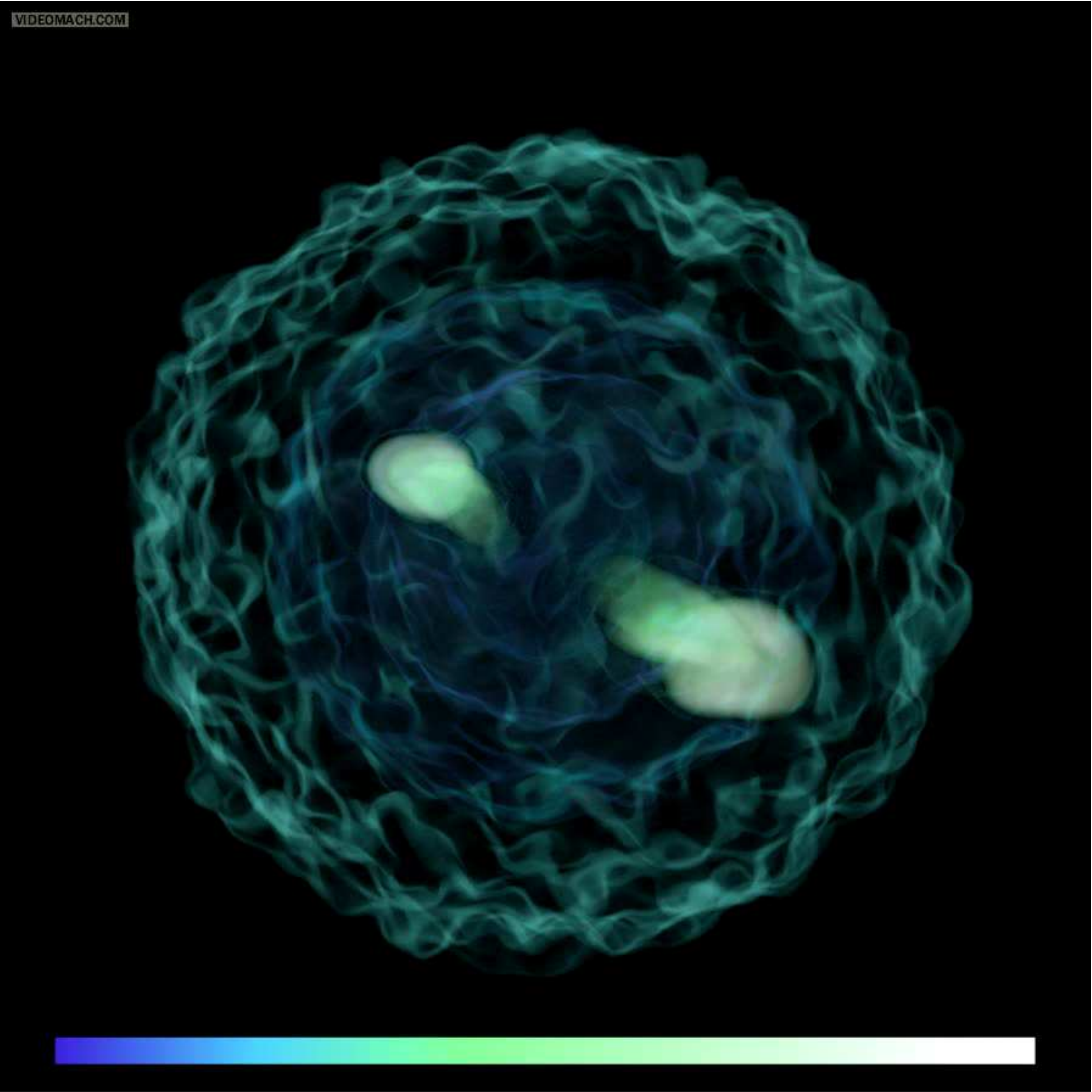}
\caption[Volume rendering of entropy for all models]{Volume rendering of entropy for the ST ({\it upper left}), I26 ({\it upper right}), I13 ({\it lower left}), and RE ({\it lower right}), after the jet disturbances have nearly reached the computational grid boundary.  Higher entropy is shown in lighter colors, and the ICM is only shown in isosurfaces.  Animations of these quantities as seen from several different angles will be available through the electronic version of this paper.}
\label{fig:ST_en}
\end{center}
\end{figure*}

Here, we present a discussion of the dynamics and morphology of our simulated flows.
As mentioned in the previous section, Figure \ref{fig:STspvr} and the associated animations clearly show the evolution of flow speed in each jet model over time from various angles.
Likewise, Figure \ref{fig:ST_en} and its associated animations show volume renderings of jet entropy surrounded by more transparent isosurface renderings of ICM entropy.
Animations such as these are typically the best way to quickly visualize causal relationships between flow structures and to convey the general character of the flow.

First, we discuss the general properties of jet length evolution as a function of time, as illustrated by Figure \ref{fig:doublejetlengths}.
In that figure, the solid lines represent the maximum distance between the cluster core and $C_j > 0.90$ in the direction of jet/relic motion and averaged over the two jets in each model.
The dotted lines show the position of the forwardmost shock, also in the direction of jet/relic motion and averaged over both jets.
To locate the shocks, we used an identification strategy similar to that employed in \citet{ryuetal03} in which the temperature and entropy gradients were of the same sign, the velocity divergence was negative, and the temperature jump was that of a Mach 1.3 shock or stronger.
In Figure \ref{fig:doublejetlengths}, we see that substantial separation develops between the bow shock and the jet material at some point in every non-steady jet model.
When the dotted line and solid line are exactly coincident, this corresponds to 
a shock with $M < 1.3$ or no shock at all.

\begin{figure*}[t]
\begin{center}
\includegraphics[type=pdf,ext=.pdf,read=.pdf,width=0.45\textwidth]{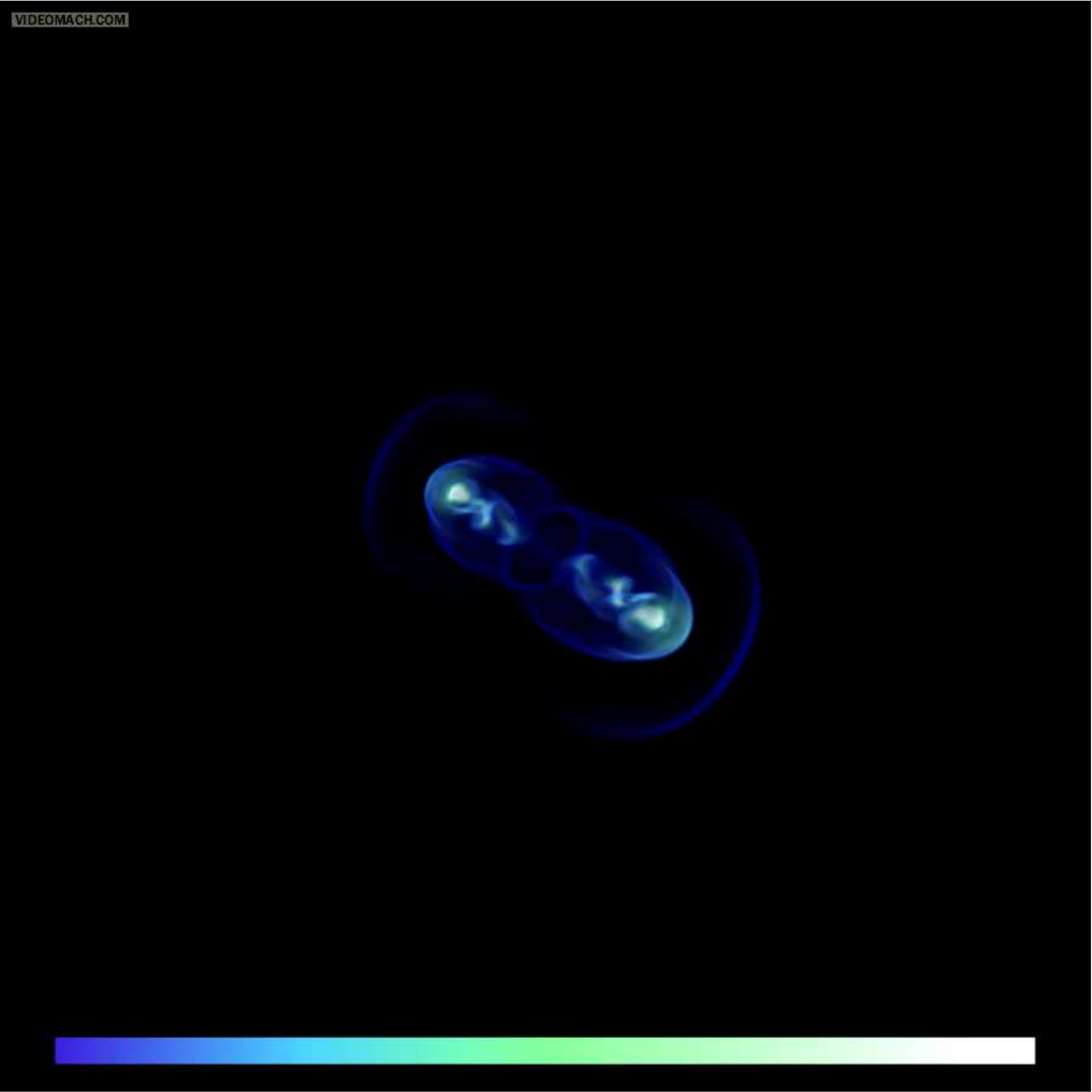}
\includegraphics[type=pdf,ext=.pdf,read=.pdf,width=0.45\textwidth]{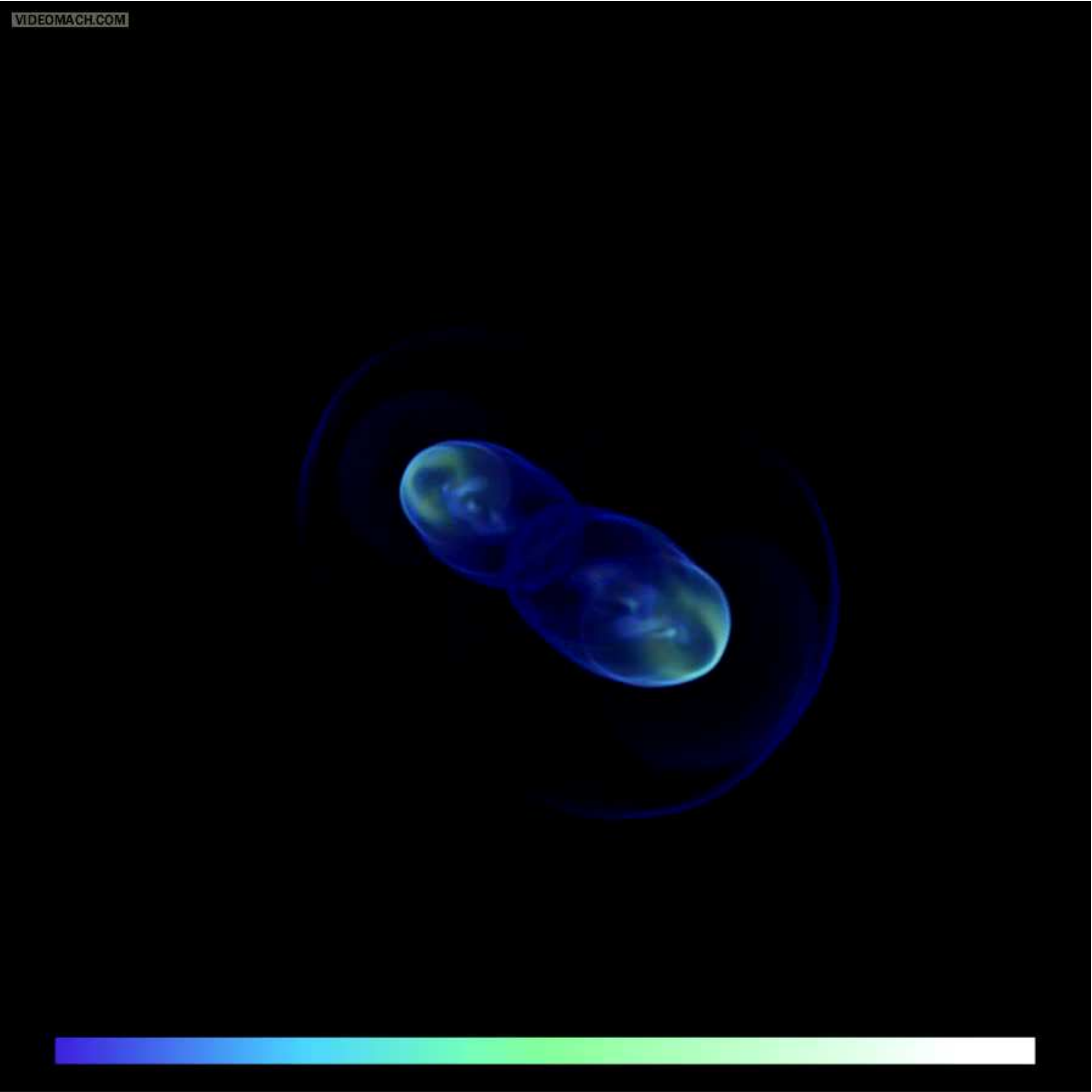}
\\[0.75em]
\includegraphics[type=pdf,ext=.pdf,read=.pdf,width=0.45\textwidth]{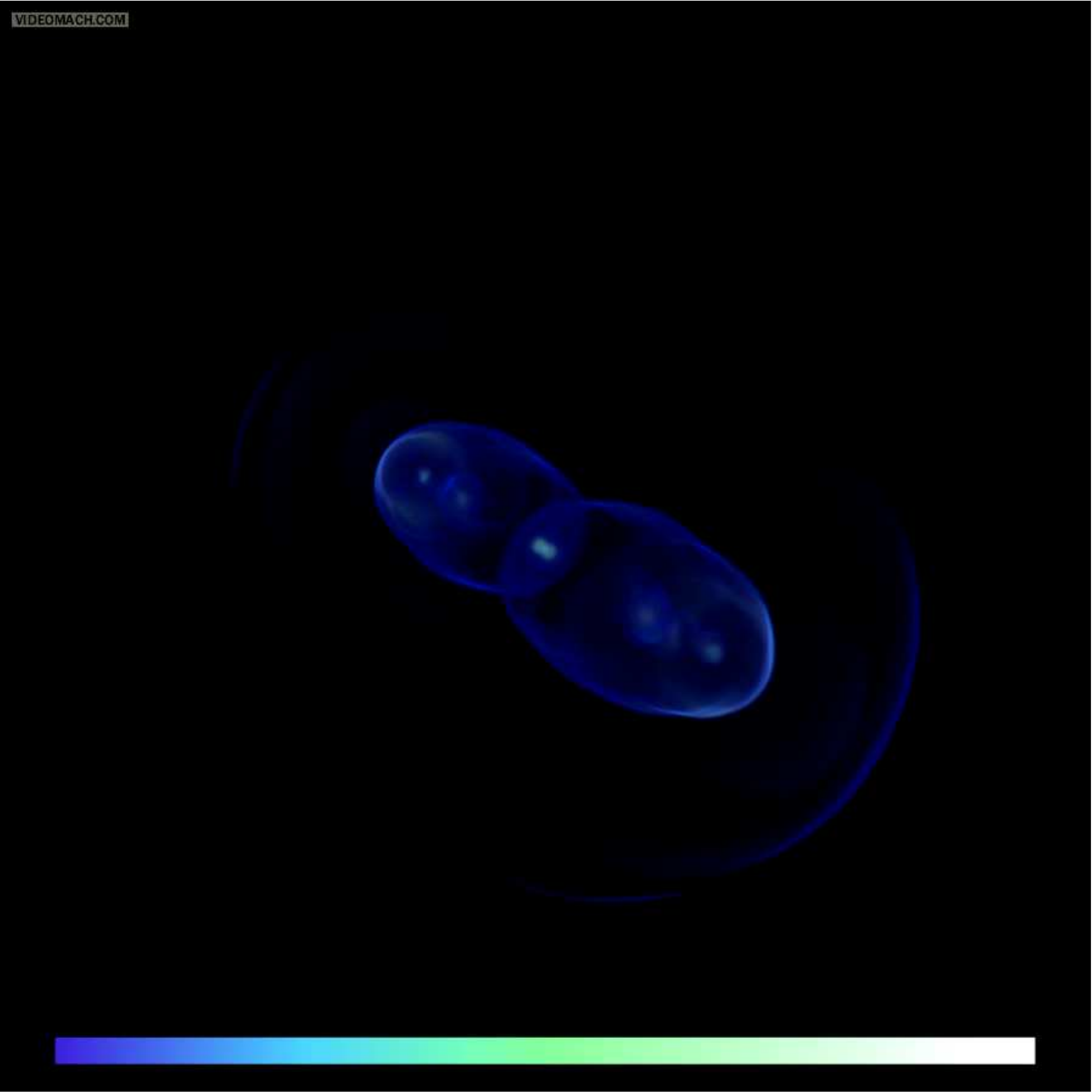}
\includegraphics[type=pdf,ext=.pdf,read=.pdf,width=0.45\textwidth]{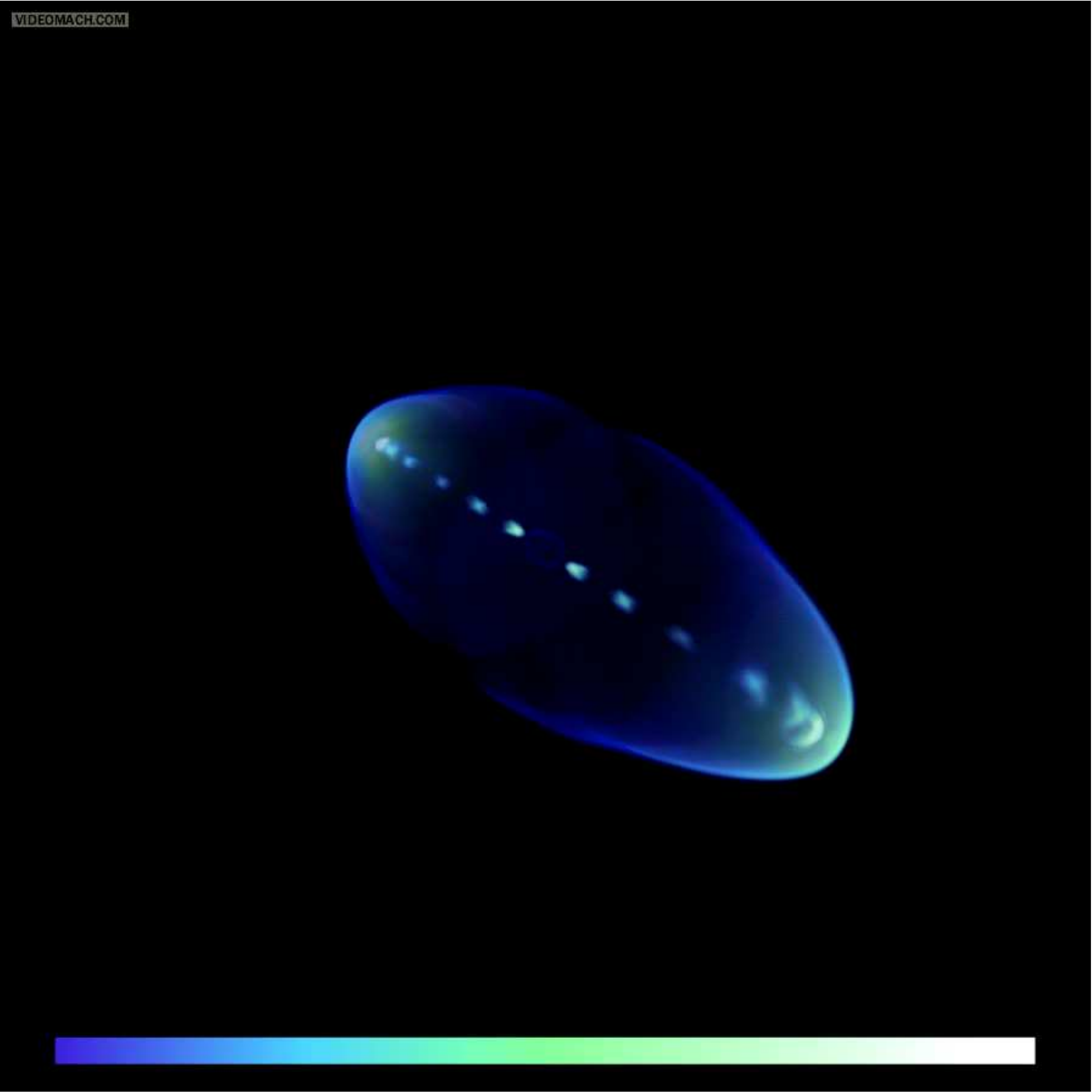}
\caption[Shock structures in the I13 model]{Shock structures in the I13 model at early ({\it upper left}), moderate ({\it upper right}), and late ({\it bottom left}) times, illustrated by volume renderings of $-{\bf \nabla \cdot v}$.  The lower-right image shows the ST model at a late time for comparison. An animation of this quantity as seen from several different angles will be available through the electronic version of this paper}
\label{fig:doublejetshocks}
\end{center}
\end{figure*}

\begin{figure*}[t]
\begin{center}
\includegraphics[type=pdf,ext=.pdf,read=.pdf,width=0.45\textwidth]{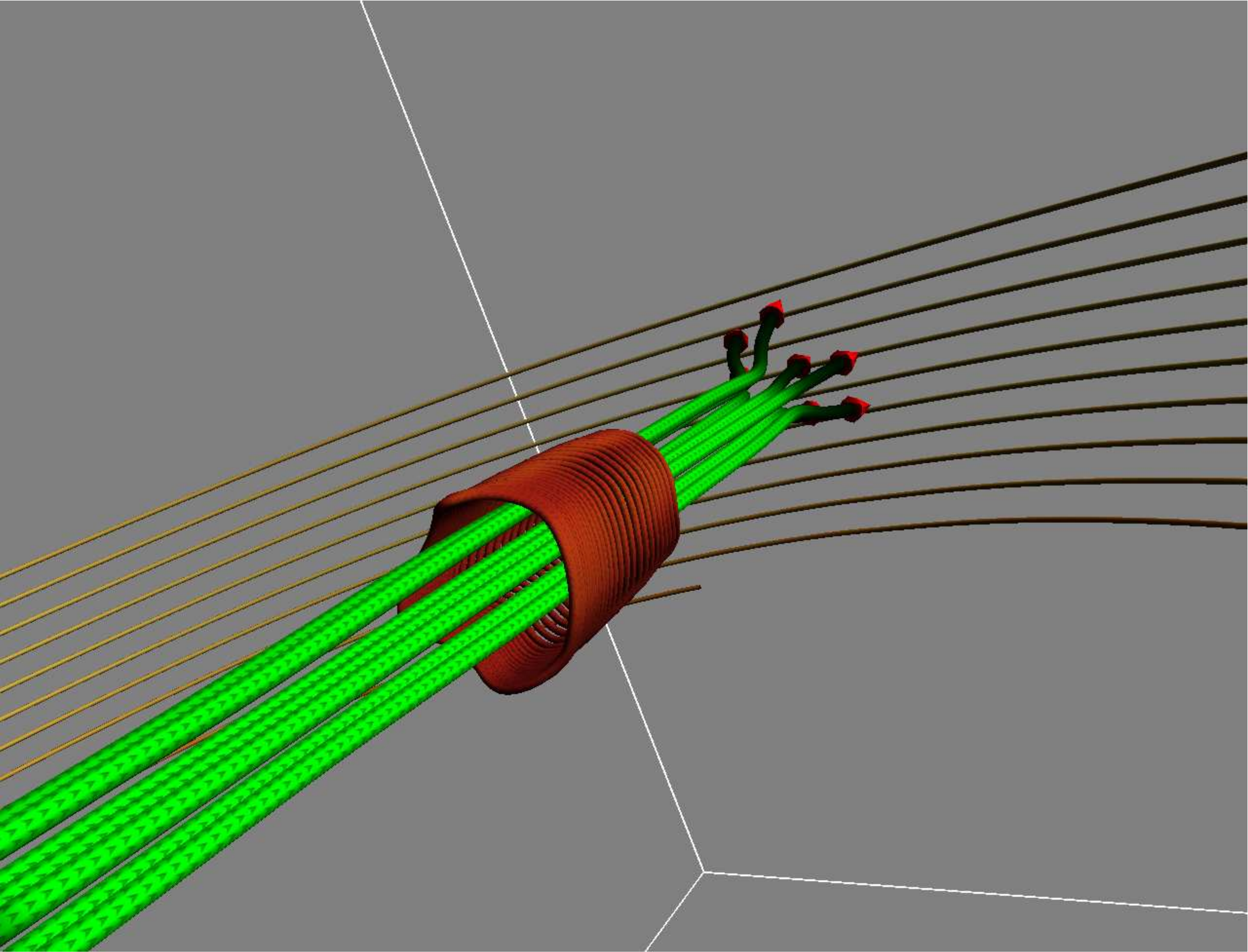}
\includegraphics[type=pdf,ext=.pdf,read=.pdf,width=0.45\textwidth]{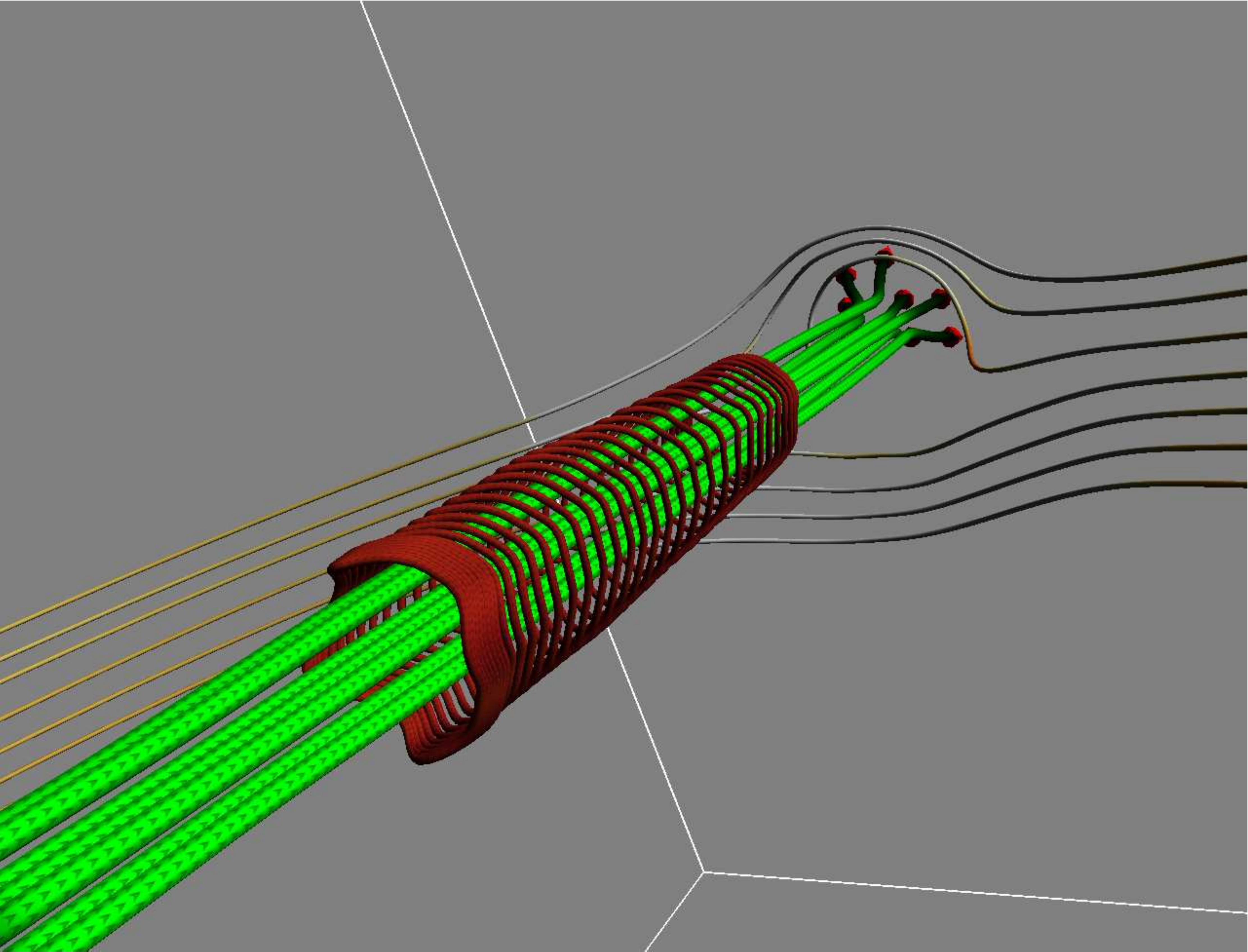}
\\[0.75em]
\includegraphics[type=pdf,ext=.pdf,read=.pdf,width=0.45\textwidth]{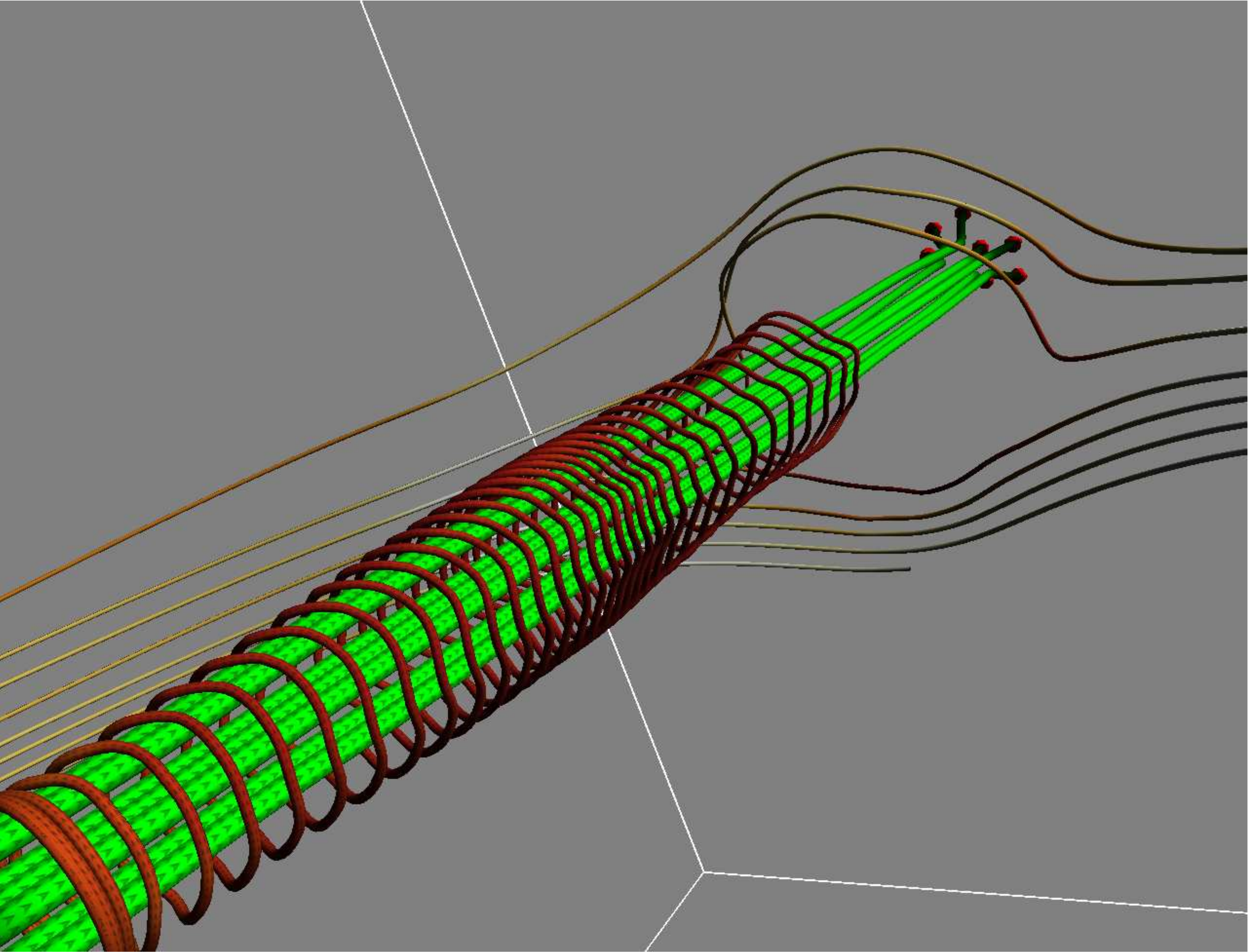}
\includegraphics[type=pdf,ext=.pdf,read=.pdf,width=0.45\textwidth]{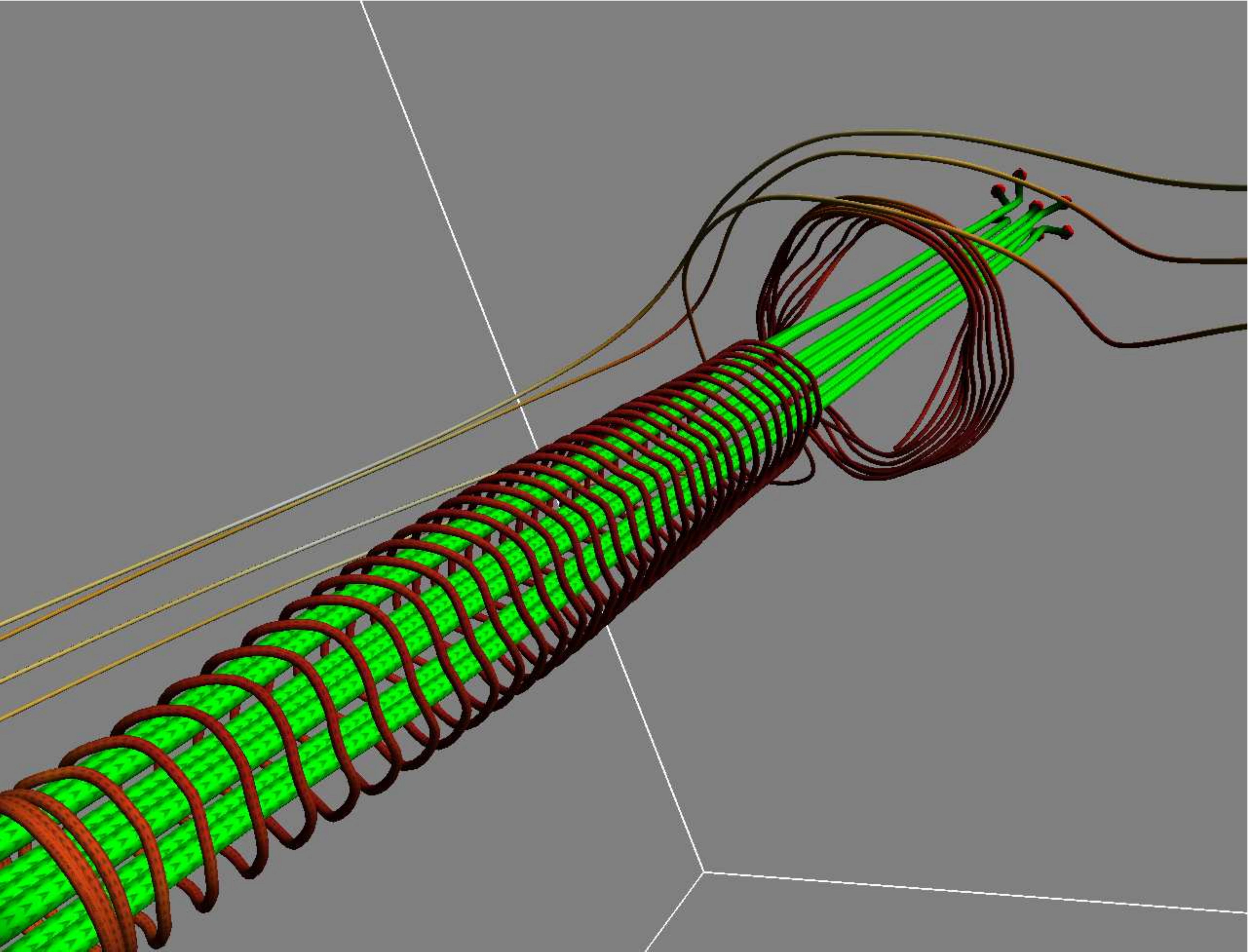}
\caption[An example of magnetic reconnection in the ST model]{Example of magnetic reconnection in the ST model.  The light-colored, straight streaklines pointing from the bottom left to the upper right of each image represent the jet velocity, which is surrounded by a toroidal magnetic field.  A set of ICM magnetic field lines sits in the path of the jet, color coded to magnetic field strength (light: strong).  From $t \sim 20~$Myr ({\it upper left}) to $t \sim 40~$Myr ({\it lower right}), the jet field pushes through and eventually connects with the ICM field structure.} 
\label{fig:STreconnection}
\end{center}
\end{figure*}

There are two interesting evolutionary features in Figure \ref{fig:doublejetlengths}.
First, we note that the ST, I26, and I13 models all asymptote to similar slopes, suggesting a common power-law evolution of $l \propto t^\delta$ for these models, where $\delta \sim 1.3$.
A simple expectation for the advance speed, gotten from self-similar scaling arguments in \citet{oneilletal05}, for example, is that
\begin{equation}
l \propto t^{\frac{3}{5-\kappa}}
\end{equation}
assuming a homologous structure for the cocoon and that the ambient density follows a power law $\rho_a \propto x^{-\kappa}$.
Although the ambient density in this case is not a strict power law, we can estimate $\kappa \sim 1.3$ around $r \sim 100~$kpc and $\kappa \sim 1.7$ around $r \sim 300~$kpc.
These values produce an expected $\delta \sim 0.8-0.9$, which is substantially flatter than the measured value of $\delta \sim 1.3$.
Self-similarity is thus probably not a reasonable assumption for an atmosphere whose density dependence changes slope so dramatically.
That the measured values of $\delta$ for the I26 and I13 models are slightly flatter than the ST model is not surprising, however, since the jet momentum is probably distributed over a slightly larger area, as we will see momentarily.

In contrast to systems in which jets actively drive the flow, the advance speed of the RE model asymptotes to a power law of approximately $\delta \sim 0.4$ as it decelerates in the ICM.
Although the residual jet flows terminate quite rapidly after being shut off, we expect the overall momentum of the system to be distributed in a mix of cocoon material and ICM that has been entrained, dragged, or otherwise pushed into motion.
We can estimate the rate at which the plumes are decelerated, assuming that the dynamics are dominated by the effects of drag resulting from motion of the relics through the ICM.
The drag force is given by $F_d \sim - C_d A \rho_a v^2$, where $C_d \sim 1$ is the drag coefficient, $A$ is the cross-sectional area of the plume, and $v$ is the relic velocity.
We assume, as before, that the ambient density follows a power law, this time specifying it more fully as $\rho_a = \rho_{a, {\rm off}} (x/x_{\rm off})^{-\kappa}$, where the ``off'' subscript refers to the value of that quantity at the position of the jet head when the jets are shut off.
If we assume that the radial expansion is adiabatic, then $PV^{\gamma} \propto P r^{3\gamma} $ is conserved, assuming the volume of the relic is proportional to the cube of its radius.
If the ambient pressure profile is proportional to that of the density (\ie the ICM is approximately isothermal), then $r \propto x^{\kappa/3\gamma}$, meaning that the relic area can be approximated as $A \approx \pi r^2 = \pi r_{\rm off}^2(x/x_{\rm off})^{2\kappa/3\gamma}$.
Under these assumptions, the relic deceleration resulting from drag can be written as
\begin{equation}
{\ddot x} \sim \Psi x^{(2\kappa/3\gamma - \kappa)} {\dot x}^2,
\end{equation}
where, writing the total mass of the relic as $m_{\rm re}$ and assuming $C_d =1$,
\begin{equation}      
\Psi = - \pi \rho_{\rm a, off} r_{\rm off}^2 x_{\rm off}^{(\kappa-2\kappa/3\gamma)}/m_{\rm re} = {\rm constant}.
\end{equation}
These expressions can be simplified considerably when we realize that, measured at $x_{\rm off} \ge 150~$kpc, $\kappa \approx 1.6-1.7 \approx \gamma$.
This reduces the drag deceleration expression to ${\ddot x} \sim \Psi x^{-1} {\dot x}^2$, which admits a solution $x \propto t^{1/{(1-\Psi)}}$.
To estimate the value of $\Psi$, we can model the initial shape of the relic plume as a cylinder of mass $m_{\rm re} = \rho_{\rm re} (\pi r_{\rm off}^2 x_{\rm off})$.
Again using the fact that $\kappa \approx \gamma$, this allows us to write $\Psi \sim - \rho_{a, {\rm off}}/\rho_{\rm re}$.
Unfortunately, using the passive color tracer, $C_j$, to estimate the relic density would fail to account for the any contribution to the total mass from the ICM component of the relic.
To more accurately estimate the relic density, we instead measure from our simulated data the average density of material moving upward with velocities above a certain threshold and positions between the jet origin and the top of the relic (thus excluding the bow shock).
Choosing only material with $v_x \ge 1.5 c_0$, which would correspond everywhere to initially supersonic motion in the unperturbed ICM, we measure an average relic density $\rho_{\rm re} \sim 0.15$ at $t \approx 26$ Myr, just after the jets are shut off.
At that relic height ($x_{\rm off} \sim 155$ kpc), $\rho_{a, {\rm off}} \sim 0.20$, resulting in an expected value of $\Psi \sim - 1.3$.
Since $l \propto t^{1/(1-\Psi)}$, the expected value of  $\delta$ is $1/(1-\Psi) \sim 0.43$, which is similar to the observed $\delta \sim 0.4$.
The closeness of the predicted and observed power laws indicates that this deceleration model is plausible, although one also assumes that this scaling would eventually break down when the relic has lost enough momentum that buoyancy becomes the dominant force in the system.

Moving on to a discussion of jet and flow morphology, we first describe steady jets as exemplified by the ST model.
Our steady jets are narrow, stable structures, a common feature in such simulations.
Since we do not perturb them at the source, but instead allow the ICM substructure to seed instabilities, they do not develop complicated shock structures like those observed in \citet{tregillisetal01b} and \citet{oneilletal05}, and the jets are not broken apart, as in \citet{heinzetal06}.
Furthermore, the jet magnetic field is sufficiently weak that the jet does not become unstable to pinch instabilities, nor does it develop an obviously modified nose cone as has been seen in some strongly magnetized jet simulations (\eg \citealt{clarkeetal86}).
As shown in the lower-right panel of Figure \ref{fig:doublejetshocks}, the bow shock is a relatively smooth surface and the internal shock structures are regularly spaced.
From the animations and image of the flow speed associated with Figure \ref{fig:STspvr}, we see that the jets are not completely axisymmetric, however, and these slight asymmetries become much more pronounced when the AGNs are shut down.
The cocoons of shocked jet material in steady jets are relatively thin and do not develop a strong backflow.
The cocoon structure is seen clearly in a volume rendering of the entropy shown in Figure \ref{fig:ST_en}, where the low-density plasma is very bright.
The ICM pressure gradient in these models is sufficient to prevent much material from traveling very far opposite the direction of jet propagation.

The I26 and I13 intermittent model jets feature less axisymmetry as the flows evolve.
Figure \ref{fig:STspvr} shows that portions of the flow with moderate speeds at late times tend to bend in response to the ICM substructure.
Likewise, we can see filaments of low-speed material that form in the cocoon.
The shock structures in these two models are very different from the steady case.
Figure \ref{fig:doublejetshocks} shows a volume rendering of the I13 shock structures at three different times.
Not only do the I13 shocks look different from those in the ST model, but they look different at a given time from earlier or later shock structures in the same model.
The animation associated with Figure \ref{fig:doublejetshocks} clearly shows the propagation of internal shocks and waves associated with restarted jet activity.
Typically, the subsequent jets form a bow shock in the old jet material that quickly advances forward through the low-density channel.
This was seen even in the earliest models of restarting jets conducted by \citet{clarkeburns91}.
A set of irregularly shaped reverse shocks that dampen into waves moves back through this material each time the jet turns off.

Compared to the ST model, the two intermittent models feature much wider cocoons that are `puffed out' by subsequent episodes of jet activity.
Figure \ref{fig:ST_en} shows images of the entropy for these models.
In the I26 model, we can see that the second episode of jet activity has run into the plume from the first episode, causing it to widen.
Likewise, the I13 model cocoon has several visible width enhancements associated with subsequent jet activity.
These features are associated in time with the measured grid increases in thermal energy, representing energy deposition in the plumes.
Interestingly, these structures persist after they are formed, presumably because subsequent jets are easily redirected by the ICM into the relatively low-density channels formed by previous jets.
The animations of entropy for the I13 model in particular show these features lasting to the end of the simulation.
There are at least two reasons, both having to do with where shocks form, that intermittent jets produce wider cocoons than their steady counterparts.
The first is that the ICM is drawn in behind the wake of the rising plume before the next intermittent jet launches.
This refilling is incomplete, as seen in the animations associated with Figure \ref{fig:ST_en}, but is sufficient to cause the next episode of jet activity to produce a shock at the base of the plume.
The second reason is that the restarted jet forms a terminal shock and multiple reverse shocks of its own (as seen in Figure \ref{fig:doublejetshocks}) as it propagates through the existing plume.
Both of these episodes of shock formation deposit thermal energy in the plume and cause the cocoon to expand laterally as a result.

Finally, the relic model is distinct from all other models in that the jet plumes completely detach from the source.
Figures \ref{fig:STspvr} and \ref{fig:ST_en} show the evolution of this model, in which two plumes of material rise in the ICM as the deposited momentum is depleted.
When the jets shut down, the flows are already well over $100~$kpc from the cluster core, and their cocoons are prolate in shape.
As the ends of the plumes decelerate, however, the relic bubbles grow less prolate. 
Since observed relic and ghost bubbles are typically seen to be intact structures, we note that the entropy at late times shows relatively rounded morphology, particularly near the leading edge of the bubble.
Significantly, we do not observe the formation of a torus, in contrast with simulations of bubbles initialized as spherical cavities in clusters that would otherwise be in hydrostatic equilibrium (see \citealt{oneilletal09} for an extensive discussion).
Finally, it is important to note that we have only explored a small part of the available parameter space with this particular relic model.
To fully evaluate whether jet/ICM magnetic fields stabilize jet-blown relic bubbles, as has been suggested by \citet{deyoung2003}, for example, further variation of magnetic field strength and geometry would be required.

\subsection{Magnetic Field Evolution}\label{sec:b}

Finally, we devote some space to a discussion of magnetic field structures in these flows.
Section \ref{sec:energy} addressed the evolution of magnetic energy in these systems, but the details of field structure evolution were not extensively discussed.
Employing a visualization package called FieldVis \citep{fieldetal07}, we are able to graphically represent magnetic field lines and to map useful flow properties on them as needed.
This is a tremendously powerful approach since the challenges of visualizing multiple three-dimensional vector and scalar fields can otherwise be overwhelming.

Figure \ref{fig:STreconnection} shows the evolution of magnetic field structures in the ST model.
At early times ({\it upper-left image}), the jet is propagating toward an undisturbed ICM field structure, similar to one of the structures seen in Figure \ref{fig:fv_fieldt0}.
In that image, the jet moves from the lower left toward the upper right of the image and is represented by a set of lightly colored velocity streaklines.
The jet is surrounded by a toroidal field, represented by the darker set of coiled lines.
As the jet disturbance first reaches the ICM ({\it upper-right image}), the ICM field lines are pushed aside and stretched.
The intensity of the magnetic field lines represents the field strength (brighter=stronger), and we see that the field is enhanced as it wraps around the jet.
This process continues in the next frame ({\it lower left}), where the jet has completely penetrated the ICM field.
In that same image, we also note evidence of an internal shock within the jet corresponding to a concentration of field lines roughly halfway down the jet structure.
In the last image ({\it lower right}), we see that the jet field has interacted with the ambient field.
To explain  this image completely, we must first point out that the identified ICM field lines were selected to lie ahead of the jet so that the jet field would necessarily intersect them.
This final image shows that at least one of the ICM field lines has become attached to the toroidal jet field.
This is seen most easily by noting that there are four visible lines to the left of the jet, but only three emerge to the right while the other is attached to the larger toroidal structure near the end of the jet.
We examined the values of the passive variable in this region to find that in fact the large toroidal field is associated with jet material and connects with a line that is associated with the ICM.
Although reconnection in our numerical scheme does not represent a significant amount of direct
energy exchange, the geometries that develop from it have the potential to affect the system energetics.
As this system evolves, the reconnected field line will get stretched out with the jet flow, possibly leading to an alignment between the jet and field orientation.
This radial realignment of ICM fields could have important implications for cluster heating via thermal conduction along field lines, as has been suggested by \citet{bogdanovicetal09} and \citet{guooh09}, for example.
Additionally, this alignment could be similar to that inferred from polarization maps of some stronger sources \citep{bridleetal94}.
We list this outcome only as a speculative possibility, however.
Constructing synthetic radio observations is the only rigorous way to determine the polarization properties of a simulated structure, and we defer that discussion to a later work.

\begin{figure}
\begin{center}
\includegraphics[type=pdf,ext=.pdf,read=.pdf,width=0.48\textwidth]{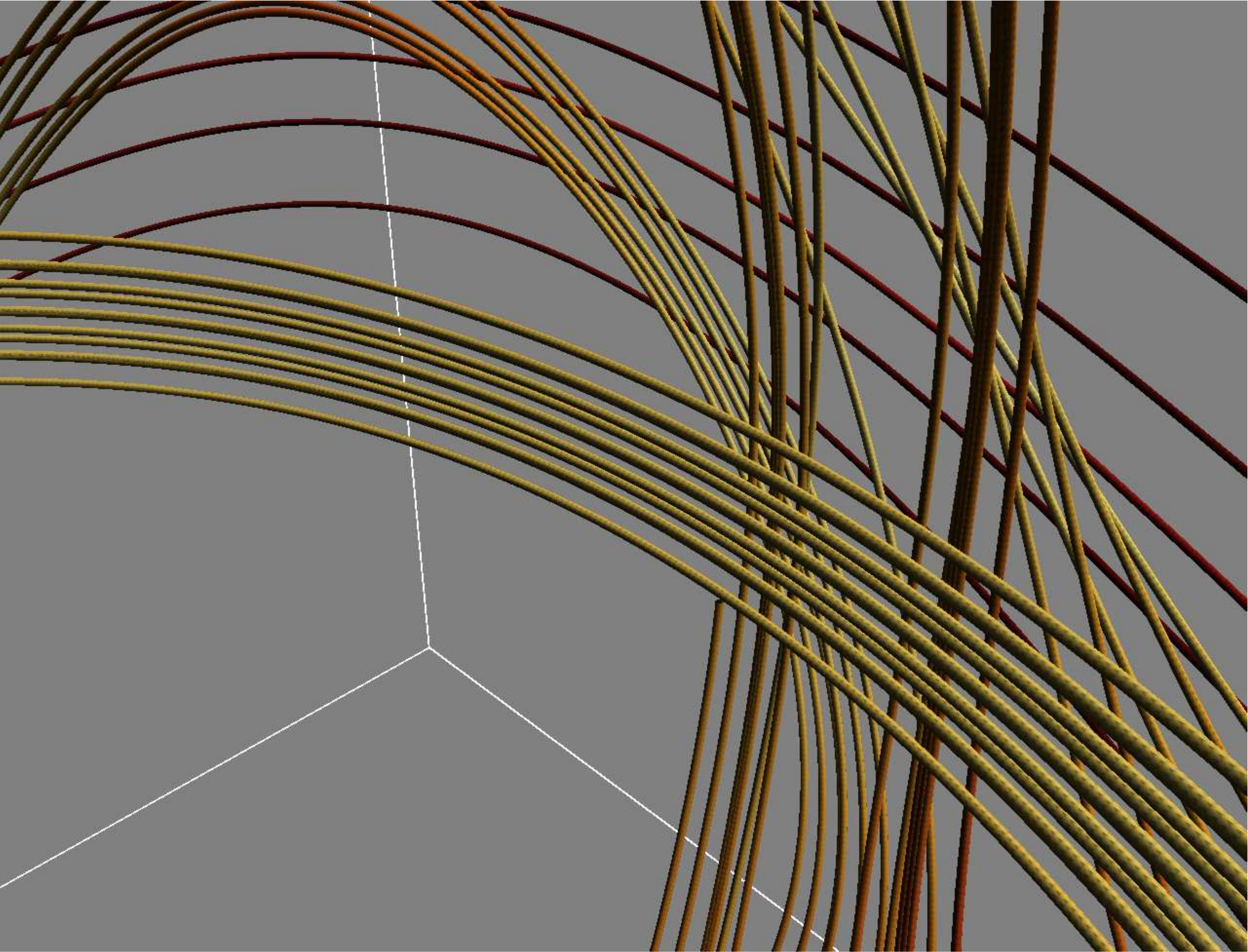}
\\[0.75em]
\includegraphics[type=pdf,ext=.pdf,read=.pdf,width=0.48\textwidth]{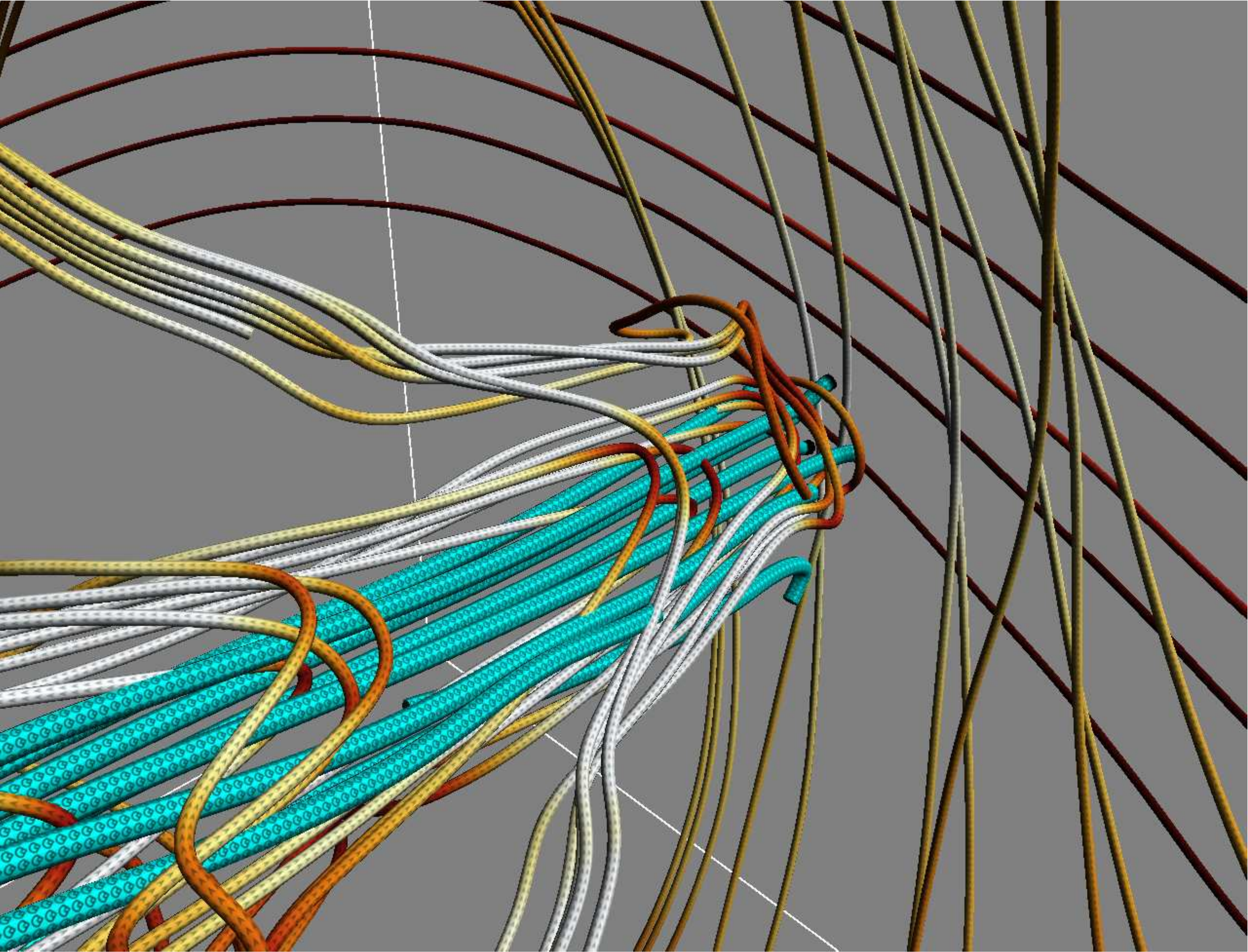}
\caption[Sample magnetic field evolution in the RE model]{Sample magnetic field evolution in the RE model.  {\it Upper panel:} the initial ICM magnetic field configuration at $r\sim 200~$kpc from the cluster core.  Intensity reflects field strength (light=strong).  {\it Lower panel:} the same region at $t\sim120~$Myr.  Now, the relic velocity is shown as streaklines with `baseball diamond' glyphs pointing in the direction of motion.  The surrounding stretched lines represent ICM field lines that have gotten stretched and amplified by the motion of the relic.}
\label{fig:relicreconnection}
\end{center}
\end{figure}

Interestingly, the ICM magnetic field seems to interact strongly with the behavior of the intermittent jets.
Although the nominal mean magnetic pressure is a factor $\beta^{-1} = 1/100$ smaller than the gas pressure, local magnetic pressure gradients can still exert a substantial force on the flow.
There are two reasons for this. First, the variations in field strength can lead to locally much smaller values of $\beta$. Second, it is important to remember that the dynamical
consequence of magnetic pressure comes through its gradient, not its magnitude, per se. When
the local length scale of magnetic field variations is smaller than the length scale
of the gas pressure variations, the dynamical role of the magnetic pressure is
stronger than the $\beta$ parameter implies.
In the ST model, these forces are swamped by the momentum of the jets and are not easily seen.
In the intermittent cases, however, there is some evidence that the ICM field influences the cocoon shape as the jets propagate.
Figure \ref{fig:ST_bm} and the associated animation illustrate this point well.
As the jet restarts, the bright regions of enhanced field in the jet coincide with the ICM enhancements.
Likewise, we can see that the cocoon width responds to these fields.
Since this is more difficult to see in other models, however, we conclude that it is the interaction between intermittency and the ambient field that makes these structures stand out.

Figure \ref{fig:relicreconnection} shows another field geometry, this time for the RE model.
The upper image shows a set of undisturbed ICM field lines at roughly $200~$kpc from the cluster core.
These have a structure much like those seen in Figure \ref{fig:fv_fieldt0} and the upper-left panel of Figure \ref{fig:STreconnection}.
Again, field strength is shown as a surface property of the lines and the structure of the constant-$\beta$ atmosphere can be seen from the darker background fields.
The lower image shows the interaction of the relic bubble and the ICM field, where the relic is represented by velocity streaklines using a different glyph.
As the relic moves from the lower left to upper right of the image, it drags a significant amount of magnetic field with it.
Specifically, we see several very strong ({\it i.e.}, brightly colored) field structures that are stretched in the direction of relic propagation.
These are ICM fields that connected with the relic bubble fields as it was rising.
Once this happens, the continued propagation of the relic serves to stretch and amplify these fields.
Again, the passive variable confirms this explanation by showing that several of the strong field lines in the lower image are associated with both jet and ICM material.
This is a prime example of how magnetic fields can get enhanced and transported in a cluster even when the central AGN engine is off.

\section{Conclusions and Astrophysical Implications}\label{sec:conc}
We have conducted a series of simulations designed to examine the role of intermittency and cluster substructure on the energetics, dynamics, and morphologies of jet-driven flows.

The most important results from this work are summarized here as follows.

1. All models of jet behavior, including steady, intermittent, and relic flows, are at least $\sim 60\%$ efficient at delivering thermal energy to their environments.
While this efficiency is remarkable, none of the jet models are particularly adept at distributing this energy uniformly throughout their environments.
Specifically, the transfer of energy from steady jets to the ICM takes place very near the jet cocoon, while intermittent jets spread the energy in only a slightly wider pattern.
Jet-blown relic bubbles are the most efficient at transferring what energy they have, but they also transmit less total energy than any other model.
Taken together, these results confirm that the lack of uniform heating in hydrodynamic systems described by \citet{vernaleoreynolds06} is also a problem for MHD jets in realistic magnetized clusters.
It is possible that the subgrid modeling of turbulence undertaken by \citet{bruggen09} could help to alleviate this problem in magnetized systems, too, but this will require a modification of the subgrid model that has been used thus far.

2. The morphologies of intermittent jets retain evidence of previous episodes of jet activity.
Specifically, subsequent bouts of jet activity encounter previously formed structures, creating reverse shocks at their interfaces.
The cocoons formed by such flows also evolve to be characteristically wider than those of steady jets. 
These structures can be associated with particular depositions of energy into the ICM and may indicate where heating is taking place.
A full exploration of whether such features are observationally detectable is deferred to
the analysis of synthetic observations from our simulated data \citep{mendygraletal10}.

3. The presence of cluster substructure can perturb initially axisymmetric jets and does so much more quickly when the jets are non-steady.
This is a very natural way to perturb jets in clusters that have not undergone recent mergers and which appear to be relaxed.
Still, as noted, the density and magnetic structures included in these simulations are small and insufficient to significantly isotropize the energy distribution in cluster cores.

4. Magnetic structures in the ICM are affected by reconnection and can get carried along with jet-driven and passively evolving flows.
Although the specific process of reconnection will never directly increase the magnetic energy, it can facilitate field geometries that operate in conjunction with the flow to amplify fields
through field line stretching and twisting.
Additionally, such processes have the potential to advect core ICM fields into the outer regions of clusters and could possibly orient such fields in the direction of jet motion, with potentially important implications for heat transfer via thermal conduction.
Interestingly, local ICM magnetic pressure gradients are capable of affecting these flows even when the global field strengths are considered to be insignificant.
While the magnetic energy increases associated with such interactions are minor, this is a specific way in which MHD jet flows with nominally weak fields differ from their hydrodynamic counterparts.

5. The motion of relic plumes produced by AGN jets is initially determined by a combination of the momentum resident in these systems and the drag force induced by moving through the ICM.
In our model, this mode of evolution persisted for 100 Myr after the jet had switched off, suggesting that buoyant models of jet propagation in clusters may only be relevant at late times or for jets with less total momentum deposition.
The resulting relic morphology and evolution is a useful reminder of the important differences between simulations of jet-formed bubbles and bubbles that are initialized as spherical cavities.

\acknowledgments
This work was supported at the University of Minnesota by NSF grant AST06-07674 and by the University of Minnesota Supercomputing Institute.  S.M.O. further acknowledges the support of a Doctoral Dissertation Fellowship granted by the University of Minnesota.  We thank Peter Mendygral for a series of helpful discussions that improved the quality of this work, and we thank the anonymous referee for their useful feedback.  Visualizations presented in this paper were constructed using the software packages FieldVis and The Hierarchical Volume Renderer (HVR), both developed at the University of Minnesota, and the commercial program Amira.

\clearpage

\appendix
\section{Appendix: Estimating the Jet Luminosity}
As noted in Section \ref{sec:jets}, the ratio of the unit kinetic to thermal luminosities contains a prefactor that is of order unity.
To see how this prefactor depends upon the details of jet injection, let us first define our jet density, pressure, and velocity profiles:

\begin{equation}
\rho_j = \rho_{j,0}~~~~~r \le r_s
\end{equation}

\begin{equation}
p_j =  p_{j,0}~~~~~r \le r_s
\end{equation}

\begin{equation}
v_j   = 
\left\{
\begin{array}{c}
\begin{aligned}
& v_{j,0} &~~~~~& r \le r_j \\
& v_{j,0}\left(\frac{r_s-r}{r_s - r_j}\right) &~~~~~& r_j < r \le r_s
\end{aligned}
\end{array}
\right.,
\end{equation}
where $r_j$ is the outer radius of the jet core, $r_s~(> r_j)$ is the outer radius of the transition sheath, and the ``zero'' subscript indicates a baseline value that could in principle be time-varying.
To understand how these profiles affect the luminosity, we first compute the kinetic luminosity, $L_k$:

\begin{equation}
\begin{aligned}
L_k &= \int_0^{r_s} \left(\frac{1}{2} \rho_j v_j^2\right) v_j dA \\
    &= \int_0^{r_j} \frac{1}{2} \rho_j v_j^3 dA + \int_{r_j}^{r_s} \frac{1}{2} \rho_j v_j^3  dA \\
    &= \pi \rho_{j,0}v_{j,0}^3 \left[\frac{r_j^2}{2} +  \int_{r_j}^{r_s} \left(\frac{r_s-r}{r_s - r_j}\right)^3r dr\right]\\
    &= \frac{\pi r_j^2}{20} \rho_{j,0} v_{j,0}^3 (\mu^2 + 3\mu + 6);~~~~~\mu \equiv r_s/r_j.
\end{aligned}
\end{equation}
Following a similar procedure for the thermal luminosity, $L_t$:
\begin{equation}
\begin{aligned}
L_t &= \int_0^{r_s} \left[\frac{\gamma}{(\gamma-1)} P_j\right] v_j dA \\
    &= \frac{\pi r_j^2 \gamma}{3 (\gamma-1)} p_{j,0} v_{j,0} (\mu^2 + \mu + 1)  
\end{aligned}
\end{equation}

We see that each of the two luminosities has a dependence upon the size of the transition region as parameterized by $\mu$.
That the two luminosities vary differently with $\mu$ simply reflects the decision to have the velocity transition smoothly to zero, independent of the density or pressure.
To recover the prefactor found in Section \ref{sec:jets}, we take the ratio of the two luminosities:
\begin{equation}
\frac{L_k}{L_t} = \frac{3 (\gamma -1) \rho_{j,0} v_{j,0}^2 (\mu^2 + 3\mu + 6)}{20 \gamma p_{j,0} (\mu^2 + \mu + 1) }
\end{equation}

Plugging in the values from Table 1, $\rho_{j,0}=8.33\times10^{-28}$ g cm$^{-3}$, $p_{j,0}=4.00\times10^{-10}$ dyne cm$^{-2}$, and $v_{j,0} = 2.69 \times 10^9$ cm s$^{-1}$,  we obtain:
\begin{equation}
\frac{L_k}{L_t} = 0.904 \left(\frac{\mu^2 + 3\mu + 6}{\mu^2 + \mu + 1}\right)
\end{equation}
For our assumed jet profile, $\mu = r_s/r_j = 7~{\rm kpc}/3~{\rm kpc}$, so the ratio of kinetic to thermal luminosity is $L_k/L_t = 1.90$.


\begin{thebibliography}{}
\bibitem[Aloy et al.(1999)]{aloyetal99} Aloy, M.~A., Ib{\'a}{\~n}ez, J.~M., Mart{\'{\i}}, J.~M., G{\'o}mez, J.-L., M{\"u}ller, E.\ 1999, \apjl, 523, L125 

\bibitem[Basson \& Alexander(2003)]{bassonalexander03} Basson, J. F., \& Alexander, P. 2003, \mnras, 339, 353

\bibitem[B\^irzan \etal(2009)]{birzanetal09} B\^irzan, L., Rafferty, D. A., McNamara, B. R., Nulsen, P. E. J., \& Wise, M. W. 2009, arXiv: 0909.0397v2

\bibitem[B\^irzan \etal(2004)]{birzanetal04} B\^irzan, L., Rafferty, D. A., McNamara, B. R., Wise, M. W., \& Nulsen, P. E. J. 2004, \apj, 607, 800

\bibitem[Blanton \etal(2009)]{blantonetal09} Blanton, E. L., Randall, S. W., Douglass, E. M., Sarazin, C. L., Clarke, T. E., \& McNamara, B. R. 2009, \apj, 697L, 95B

\bibitem[Bogdanovi\'c \etal(2009)]{bogdanovicetal09} Bogdanovi\'c, T., Reynolds, C. S., Balbus, S. A., \& Parrish, I. J. 2009, \apj, 704, 211

\bibitem[Bridle \etal(1994)]{bridleetal94} Bridle, A. H., Hough, D. H., Lonsdale, C. J., Burns, J. O., Laing, R. A. 1994, \aj, 108, 3

\bibitem[Br\"uggen \etal(2005)]{bruggenetal05} Br\"uggen, M., Ruszkowski, M., \& Hallman, E. 2005, \apj, 630, 740

\bibitem[Br\"uggen \& Scannapieco(2009)]{bruggen09} Br\"uggen, M. \& Scannapieco, E. 2009, \mnras, 398, 548

\bibitem[Br\"uggen \etal(2009)]{bruggenetal09} Br\"uggen, M., Scannapieco, E., \& Heinz, S. 2009, \mnras, 395, 2210

\bibitem[Carvalho \& O'Dea(2002a)]{carvalhoodea02a} Carvalho, J. C., \& O'Dea, C. P. 2002, \apjs, 141, 337

\bibitem[Carvalho \& O'Dea(2002b)]{carvalhoodea02b} Carvalho, J. C., \& O'Dea, C. P. 2002, \apjs, 141, 371

\bibitem[Cioffi \& Blondin(1992)]{cioffiblondin92} Cioffi, D. F., \& Blondin, J. M. 1992, \apj, 392, 458

\bibitem[Clarke et al.(1986)]{clarkeetal86} Clarke, D.~A., Norman, M.~L., \& Burns, J.~O.\ 1986, \apjl, 311, L63 

\bibitem[Clarke \& Burns(1991)]{clarkeburns91} Clarke, D. A., \& Burns, J. O. 1991, \apj, 369, 308

\bibitem[Croom \etal(2004)]{croometal04} Croom, S. M., Boyle, B. J., Shanks, T., Outram, P., Smith, R. J., Miller, L., Loaring, N., Kenyon, S., \& Couch, W. 2004, AGN Physics with the Sloan Digital Sky Survey, ASP Conference Series, eds. G. T. Richards \& P. B. Hall, 311, 457

\bibitem[De Young(2003)]{deyoung2003} De Young, D.~S.\ 2003, \mnras, 343, 719 

\bibitem[Diehl \etal(2008)]{diehletal08} Diehl, S., Li, H., Fryer, C. L., \& Rafferty, D. 2008, \apj, 687, 173

\bibitem[Dong \& Stone(2009)]{dongstone09} Dong, R., \& Stone, J. M. 2009, \apj, 704, 1309

\bibitem[Duncan \& Hughes(1994)]{duncanhughes94} Duncan, G.~C., \& Hughes, P.~A.\ 1994, \apjl, 436, L119 

\bibitem[Dunn \etal(2005)]{dunnetal05} Dunn, R. J. H., Fabian, A. C., \& Taylor, G. B. 2005, \mnras, 364, 1343

\bibitem[Ettori \etal(1998)]{ettorietal98} Ettori, S., Fabian, A. C., White, D. A., 1998, \mnras, 300, 837

\bibitem[Fabian \etal(2000)]{fabianetal00} Fabian, A. C., Sanders, J. S., Ettori, S., Taylor, G. B., Allen, S. W., Crawford, C. S., Iwasawa, K., Johnstone, R. M., Ogle, P. M. 2000, \mnras, 318L, 65F

\bibitem[Fabian \etal(2001)]{fabianetal01} Fabian, A. C., Mushotzky, R. F., Nulsen, P. E. J., \& Peterson, J. R. \mnras, 321, L20

\bibitem[Fabian \etal(2003)]{fabianetal03}
Fabian, A. C., Sanders, J. S., Allen, S. W., Crawford, C. S., Iwasawa, K., Johnstone, R. M., Schmidt, R. W., \& Taylor, G. B. 2003, \mnras, 344, L43

\bibitem[Field \etal(2007)]{fieldetal07} Field, B., O'Neill, S., Urness, T., Interrante, V., \& Jones, T. W. 2007, IEEE Comput. Graph. Appl., 27(1), 9

\bibitem[Forman \etal(2005)]{formanetal05} Forman, W., Nulsen, P., Heinz, S., Owen, F., Eilek, J., Vikhlinin, A., Markevitch, M., Kraft, R., Churazov, E., \& Jones, C. 2005, \apj, 635, 894

\bibitem[Fujita \etal(2002)]{fujitaetal02} Fujita, Y., Sarazin, C. L., Kempner, J. C., Rudnick, L., Slee, O. B., Roy, A. L., Andernach, H., Ehle, M. 2002, \apj, 575, 764

\bibitem[Gaibler et al.(2009)]{gaibleretal09} Gaibler, V., Krause, M., \& Camenzind, M.\ 2009, \mnras, 1611 

\bibitem[Guo \& Oh(2009)]{guooh09} Guo, F. \& Oh, S. P. 2009, \mnras, in press

\bibitem[Hardee \& Clarke(1992)]{hardeeclarke92} Hardee, P. E., \& Clarke, D. A. 1992, \apj, 400, L9

\bibitem[Heinz \etal(2006)]{heinzetal06} Heinz, S., Br\"uggen, M., Young, A., \& Levesque, E. 2006, \mnras, 373, L65

\bibitem[Heinz \& Churazov(2005)]{heinzchurazov05} Heinz, S., \& Churazov, E. 2005, \apj, 634, L141

\bibitem[Jamrozy \etal(2007)]{jamrozyetal07} Jamrozy, M., Konar, C., Saikia, D. J., Stawarz, L., Mack, K.-H., \& Siemiginowska, A. 2007, \mnras, 378, 581

\bibitem[Jones \& DeYoung(2005)]{jonesdeyoung05} Jones, T. W. \& DeYoung, D. S. 2005, \apj, 624, 586

\bibitem[Jones \etal(1999)]{jonesetal99} Jones, T. W., Ryu, D., \& Engel, A. 1999, \apj, 512, 105

\bibitem[Kaastra \etal(2001)]{kaastraetal01} Kaastra, J. S., Ferrigno, C., Tamura, T., Paerels, F. B. S., Peterson, J. R., \& Mittaz, J. P. D. 2001, \aap, 365L, 99K

\bibitem[Kaiser \& Alexander(1997)]{kaiseralexander97} Kaiser, C. R., \& Alexander, P. 1997, \mnras, 286, 215

\bibitem[Keppens \etal(2008)]{keppensetal08} Keppens, R., Meliani, Z., van der Holst, B., \& Casse, F.\ 2008, \aap, 486, 663 

\bibitem[Kolmogorov(1941)]{kolmogorov41} Kolmogorov, A. N. 1941, Proc. R. Soc. A, 434, 9

\bibitem[Komissarov \& Falle(1998)]{komissarovfalle98} Komissarov, S. S., \& Falle, S. A. E. G. 1998, \mnras, 297, 1087

\bibitem[Krause(2003)]{krause03} Krause, M. 2003, \aap, 398, 113

\bibitem[Krause(2005)]{krause05} Krause, M. 2005, \aap, 431, 45

\bibitem[Marti et al.(1997)]{martietal97} Marti, J.~M.~A., Mueller, 
E., Font, J.~A., Ibanez, J.~M.~A., \& Marquina, A.\ 1997, \apj, 479, 151 

\bibitem[Mazzotta \etal(2002)]{mazzottaetal02} Mazzotta, P., Kaastra, J. S., Paerels, F. B., Ferrigno, C., Colafrancesco, S., Mewe, R., \& Forman, W. R. 2002, \apj, 567, L37

\bibitem[McNamara \etal(2000)]{mcnamaraetal00} McNamara, B. R., Wise, M., Nulsen, P. E. J., David, L. P., Sarazin, C. L., Bautz, M., Markevitch, M., Vikhlinin, A., Forman, W. R., Jones, C., \& Harris, D. E. 2000, \apj, 534L, 135M

\bibitem[McNamara \etal(2001)]{mcnamaraetal01} McNamara, B. R., Wise, M. W., Nulsen, P. E. J., David, L. P., Carilli, C. L., Sarazin, C. L., O'Dea, C. P., Houck, J., Donahue, M., Baum, S., Voit, M., O'Connell, R. W., \& Koekemoer, A. 2001, \apj, 562, L149

\bibitem[Mendygral \etal(in prep)]{mendygraletal10} Mendygral, P. J., O'Neill, S. M., Jones, T. W., in prep.

\bibitem[Mignone et al.(2009)]{mignoneetal09} Mignone, A., Rossi, P., 
Bodo, G., Ferrari, A., \& Massaglia, S.\ 2009, arXiv:0908.4523 

\bibitem[Mizuno et al.(2007)]{mizunoetal07} Mizuno, Y., Hardee, P., 
\& Nishikawa, K.-I.\ 2007, \apj, 662, 835 

\bibitem[Navarro \etal(1997)]{nfw97} Navarro, J. F., Frenk, C. S., \& White, S. D. M. 1997, \apj, 490, 493

\bibitem[Norman(1996)]{norman96} Norman, M. L. 1996, in ASP Conf. Ser. 100, Energy Transport in Radio Galaxies and Quasars, ed. P. E. Hardee, A. H. Bridle, and J. A. Zensus (San Francisco: ASP), 319

\bibitem[Norman \etal(1982)]{normanetal82} Norman, M. L., Winkler, K.-H. A., Smarr, L., \& Smith, M. D. 1982, \aap, 113, 285

\bibitem[Nulsen \etal(2002)]{nulsenetal02} Nulsen, P. E. J., David, L. P., McNamara, B. R., Jones, C., Forman, W. R., \& Wise, M. 2002, \apj, 568, 163

\bibitem[Omma \& Binney(2004)]{ommabinney04} Omma, H., \& Binney, J. 2004, \mnras, 350, L13

\bibitem[Omma \etal(2004)]{ommaetal04} Omma, H., Binney, J., Bryan, G., \& Slyz, A. 2004, \mnras, 348, 1105

\bibitem[O'Neill \etal(2009)]{oneilletal09} O'Neill, S. M., De Young, D. S., Jones, T. W. 2009, \apj, 694, 1317

\bibitem[O'Neill \etal(2005)]{oneilletal05} O'Neill, S. M., Tregillis, I. L., Jones, T. W., \& Ryu, Dongsu 2005, \apj, 633, 717

\bibitem[Ota \& Mitsuda(2004)]{otamitsuda04} Ota, N., \& Mitsuda, K. 2004, \aap, 428, 757

\bibitem[Peterson \etal(2001)]{petersonetal01} Peterson, J. R., Paerels, F. B. S., Kaastra, J. S., Arnaud, M., Reiprich, T. H., Fabian, A. C., Mushotzky, R. F., Jernigan, J. G., \& Sakelliou, I. 2001, \aap, 365, L104

\bibitem[Reynolds \etal(2002)]{reynoldsetal02} Reynolds, C. S., Heinz, S., \& Begelman, M. C. 2002, \mnras, 332, 271

\bibitem[Robinson \etal(2004)]{robinsonetal04} Robinson, K., Dursi, L. J., Ricker, P. M., Rosner, R., Calder, A. C., Zingale, M., Truran, J. W., Linde, T., Caceres, A., Fryxell, B., Olson, K., Riley, K., Siegel, A., Vladimirova, N. 2004, \apj, 601, 621

\bibitem[Rosen et al.(1999)]{rosenetal99} Rosen, A., Hughes, P.~A., 
Duncan, G.~C., \& Hardee, P.~E.\ 1999, \apj, 516, 729 

\bibitem[Ruszkowski \etal(2007)]{ruszkowskietal07} Ruszkowski, M., En \ss lin, T. A., Br\:uggen, M., Heinz, S., \& Pfrommer, C. 2007, \mnras, 378, 662R

\bibitem[Ryu \& Jones(1995)]{ryuj95} Ryu, D. \& Jones, T. W. 1995, \apj, 442, 228

\bibitem[Ryu \etal(2003)]{ryuetal03} Ryu, D., Kang, H., Hallman, E., \& Jones, T. W. 2003, \apj, 593, 599

\bibitem[Ryu \etal(1998)]{ryuetal98} Ryu, D., Miniati, F., Jones, T. W., \& Frank, A. 1998, \apj, 509, 244

\bibitem[Sanders \etal(2009)]{sandersetal09} Sanders, J. S., Fabian, A. C., \& Taylor, G. B. 2009, \mnras, 393, 71S

\bibitem[Schoenmakers \etal(2000)]{schoenmakersetal00} Schoenmakers, A. P., de Bruyn, A. G., R\"ottgering, H. J. A., van der Laan, H., \& Kaiser, C. R. 2000, \mnras, 315, 371

\bibitem[Schuecker \etal(2004)]{schueckeretal04} Schuecker, P., Finoguenov, A., Miniati, F., B\"ohringer, H., \& Briel, U. G. 2004, \aap, 426, 387

\bibitem[Sijacki \& Springel(2006)]{sijackispringel06} Sijacki, D., \& Springel, V. 2006, \mnras, 366, 397

\bibitem[Soker \etal(2001)]{sokeretal01} Soker, N., White, R. E., David, L. P., McNamara, B. R. 2001, \apj, 549, 832

\bibitem[Tamura \etal(2001)]{tamuraetal01} Tamura, T., Kaastra, J. S., Peterson, J. R., Paerels, F. B. S., Mittaz, J. P. D., Trudolyubov, S. P., Stewart, G., Fabian, A. C., Mushotzky, R. F., Lumb, D. H., \& Ikebe, Y. 2001, \aap , 365L, 87T

\bibitem[Tregillis \etal(2001a)]{tregillisetal01a} Tregillis, I. L., Jones, T. W., \& Ryu, D. 2001a in ASP Conference Series 250, Particles and Fields in Radio Galaxies, ed. R. A. Laing, K. M. Blundell (San Francisco: ASP), 324

\bibitem[Tregillis \etal(2001b)]{tregillisetal01b} Tregillis, I. L., Jones, T. W., \& Ryu, D. 2001b, \apj, 557, 475

\bibitem[Tregillis \etal(2004)]{tregillisetal04} Tregillis, I. L., Jones, T. W., \& Ryu, D. 2004, \apj, 601, 778

\bibitem[Vernaleo \& Reynolds(2006)]{vernaleoreynolds06} Vernaleo, J. C., \& Reynolds, C. S. 2006, \apj, 645, 83

\bibitem[Vikhlinin \etal(2006)]{vikhlininetal06} Vikhlinin, A., Kravtsov, A., Forman, W., Jones, C., Markevitch, M., Murray, S. S., Van Speybroeck, L. 2006, \apj, 640, 691

\bibitem[Vogt \& En\ss lin(2005)]{vogtensslin05} Vogt, C., \& En\ss lin, T. A. 2005, \aap, 434, 67

\bibitem[Wise \etal(2007)]{wiseetal07} Wise, M. W., McNamara, B. R., Nulsen, P. E. J., Houck, J. C., \& David, L. P. 2007, \apj, 659, 1153

\bibitem[Zanni \etal(2005)]{zannietal05} Zanni, C., Murante, G., Bodo, G., Massaglia, S., Rossi, P., Ferrari, A. 2005, \aap, 429, 399

\end{thebibliography}
\end{document}